\documentclass[twocolumn,showpacs,preprintnumbers,nofootinbib,prd,
superscriptaddress,10pt]{revtex4-2}

\pdfoutput=1

\usepackage[bookmarks, pdfusetitle, pdfauthor={Stefano Schmidt}]{hyperref}
\usepackage{amsmath,amssymb}
\usepackage{amsfonts}
\usepackage{mathtools}
\usepackage[normalem]{ulem}
\usepackage{textcomp}
\usepackage{enumitem}
\usepackage{bm}
\usepackage{bbm}
\usepackage{afterpage}
\usepackage{graphicx}
\graphicspath{{img/}} 
\usepackage{tensor}

\usepackage{tabularx}
\usepackage{makecell}
\usepackage{multirow}
\usepackage{booktabs}
\usepackage{orcidlink}

\usepackage{layouts}
\usepackage[utf8]{inputenc}
\usepackage{blindtext}
\usepackage{graphicx}
\usepackage{siunitx}
\renewcommand{\d}[1]{\ensuremath{\operatorname{d}\!{#1}}}

\newcommand{\scalar}[2]{\langle #1|#2 \rangle}

\newcommand{\rescalar}[2]{( #1 |#2 )}

\newcommand{\imscalar}[2]{[ #1|#2 ]}

\begin{document}

\begin{abstract}
	Precession in Binary Black Holes (BBH) is caused by the failure of the Black Hole spins to be aligned and its study can open up new perspectives in gravitational waves (GW) astronomy, providing, among other advancements, a precise measure of distance and an accurate characterization of the BBH spins. However, detecting precessing signals is a highly non-trivial task, as standard matched filtering pipelines for GW searches are built on many assumptions that do not hold in the precessing case.
	This work details the upgrades made to the GstLAL pipeline to facilitate the search for precessing BBH signals. The implemented changes in the search statistics and in the signal consistency test are then described in detail. The performance of the upgraded pipeline is evaluated through two extensive searches of precessing signals, targeting two different regions in the mass space, and the consistency of the results is examined. 	Additionally, the benefits of the upgrades are assessed by comparing the sensitive volume of the precessing searches with two corresponding traditional aligned-spin searches.
	While no significant sensitivity improvement is observed for precessing binaries with mass ratio $q\lesssim 6$, a volume increase of up to 100\% is attainable for heavily asymmetric systems with largely misaligned spins. Furthermore, our findings suggest that the primary cause of degraded performance in an aligned-spin search targeting precessing signals is not a poor signal-to-noise-ratio recovery but rather the failure of the $\xi^2$ signal-consistency test.
	Our work paves the way for a large-scale search for precessing signals, which could potentially result in exciting future detections.
\end{abstract}


 \title{Searching for gravitational-wave signals from precessing black hole binaries with the GstLAL pipeline}
	\author{Stefano Schmidt \orcidlink{0000-0002-8206-8089}}
	\email{s.schmidt@uu.nl}
	\affiliation{Nikhef, Science Park 105, 1098 XG, Amsterdam, The Netherlands}
	\affiliation{Institute for Gravitational and Subatomic Physics (GRASP),
Utrecht University, Princetonplein 1, 3584 CC Utrecht, The Netherlands}
        
	\author{Sarah Caudill \orcidlink{0000-0002-8927-6673}}
	\affiliation{Department of Physics, University of Massachusetts, Dartmouth, MA 02747, USA}
	\affiliation{Center for Scientific Computing and Data Science Research, University of Massachusetts, Dartmouth, MA 02747, USA}
    
	\author{Jolien D. E. Creighton \orcidlink{0000-0003-3600-2406}}
	\affiliation{Leonard E.\ Parker Center for Gravitation, Cosmology, and Astrophysics, University of Wisconsin-Milwaukee, Milwaukee, WI 53201, USA}

	\author{Ryan Magee \orcidlink{0000-0001-9769-531X}}
	\affiliation{LIGO Laboratory, California Institute of Technology, Pasadena, CA 91125, USA}

	\author{Leo Tsukada  \orcidlink{0000-0003-0596-5648}}
	\affiliation{Department of Physics, The Pennsylvania State University, University Park, PA 16802, USA}
	\affiliation{Institute for Gravitation and the Cosmos, The Pennsylvania State University, University Park, PA 16802, USA}

	\author{Shomik Adhicary}
	\affiliation{Department of Physics, The Pennsylvania State University, University Park, PA 16802, USA}
	\affiliation{Institute for Gravitation and the Cosmos, The Pennsylvania State University, University Park, PA 16802, USA}

	\author{Pratyusava Baral \orcidlink{0000-0001-6308-211X}}
	\affiliation{Leonard E.\ Parker Center for Gravitation, Cosmology, and Astrophysics, University of Wisconsin-Milwaukee, Milwaukee, WI 53201, USA}

	\author{Amanda Baylor \orcidlink{0000-0003-0918-0864}}
	\affiliation{Leonard E.\ Parker Center for Gravitation, Cosmology, and Astrophysics, University of Wisconsin-Milwaukee, Milwaukee, WI 53201, USA}

	\author{Kipp Cannon \orcidlink{0000-0003-4068-6572}}
	\affiliation{RESCEU, The University of Tokyo, Tokyo, 113-0033, Japan}

	\author{Bryce Cousins \orcidlink{0000-0002-7026-1340}}
	\affiliation{Department of Physics, University of Illinois, Urbana, IL 61801 USA}
	\affiliation{Department of Physics, The Pennsylvania State University, University Park, PA 16802, USA}
	\affiliation{Institute for Gravitation and the Cosmos, The Pennsylvania State University, University Park, PA 16802, USA}

	\author{Becca Ewing}
	\affiliation{Department of Physics, The Pennsylvania State University, University Park, PA 16802, USA}
	\affiliation{Institute for Gravitation and the Cosmos, The Pennsylvania State University, University Park, PA 16802, USA}

	\author{Heather Fong}
	\affiliation{Department of Physics and Astronomy, University of British Columbia, Vancouver, BC, V6T 1Z4, Canada}
	\affiliation{RESCEU, The University of Tokyo, Tokyo, 113-0033, Japan}
	\affiliation{Graduate School of Science, The University of Tokyo, Tokyo 113-0033, Japan}

	\author{Richard N. George \orcidlink{0000-0002-7797-7683}}
	\affiliation{Center for Gravitational Physics, University of Texas at Austin, Austin, TX 78712, USA}

	\author{Patrick Godwin}
	\affiliation{LIGO Laboratory, California Institute of Technology, MS 100-36, Pasadena, California 91125, USA}
	\affiliation{Department of Physics, The Pennsylvania State University, University Park, PA 16802, USA}
	\affiliation{Institute for Gravitation and the Cosmos, The Pennsylvania State University, University Park, PA 16802, USA}

	\author{Chad Hanna}
	\affiliation{Department of Physics, The Pennsylvania State University, University Park, PA 16802, USA}
	\affiliation{Institute for Gravitation and the Cosmos, The Pennsylvania State University, University Park, PA 16802, USA}
	\affiliation{Department of Astronomy and Astrophysics, The Pennsylvania State University, University Park, PA 16802, USA}
	\affiliation{Institute for Computational and Data Sciences, The Pennsylvania State University, University Park, PA 16802, USA}

	\author{Reiko Harada}
	\affiliation{RESCEU, The University of Tokyo, Tokyo, 113-0033, Japan}
	\affiliation{Graduate School of Science, The University of Tokyo, Tokyo 113-0033, Japan}

	\author{Yun-Jing Huang \orcidlink{0000-0002-2952-8429}}
	\affiliation{Department of Physics, The Pennsylvania State University, University Park, PA 16802, USA}
	\affiliation{Institute for Gravitation and the Cosmos, The Pennsylvania State University, University Park, PA 16802, USA}

	\author{Rachael Huxford}
	\affiliation{Department of Physics, The Pennsylvania State University, University Park, PA 16802, USA}
	\affiliation{Institute for Gravitation and the Cosmos, The Pennsylvania State University, University Park, PA 16802, USA}

	\author{Prathamesh Joshi \orcidlink{0000-0002-4148-4932}}
	\affiliation{Department of Physics, The Pennsylvania State University, University Park, PA 16802, USA}
	\affiliation{Institute for Gravitation and the Cosmos, The Pennsylvania State University, University Park, PA 16802, USA}

	\author{James Kennington \orcidlink{0000-0002-6899-3833}}
	\affiliation{Department of Physics, The Pennsylvania State University, University Park, PA 16802, USA}
	\affiliation{Institute for Gravitation and the Cosmos, The Pennsylvania State University, University Park, PA 16802, USA}

	\author{Soichiro Kuwahara}
	\affiliation{RESCEU, The University of Tokyo, Tokyo, 113-0033, Japan}
	\affiliation{Graduate School of Science, The University of Tokyo, Tokyo 113-0033, Japan}

	\author{Alvin K. Y. Li \orcidlink{0000-0001-6728-6523}}
	\affiliation{LIGO Laboratory, California Institute of Technology, Pasadena, CA 91125, USA}

	\author{Duncan Meacher \orcidlink{0000-0001-5882-0368}}
	\affiliation{Leonard E.\ Parker Center for Gravitation, Cosmology, and Astrophysics, University of Wisconsin-Milwaukee, Milwaukee, WI 53201, USA}

	\author{Cody Messick}
	\affiliation{MIT Kavli Institute for Astrophysics and Space Research, Massachusetts Institute of Technology, Cambridge, MA 02139, USA}

	\author{Soichiro Morisaki \orcidlink{0000-0002-8445-6747}}
	\affiliation{Institute for Cosmic Ray Research, The University of Tokyo, 5-1-5 Kashiwanoha, Kashiwa, Chiba 277-8582, Japan}
	\affiliation{Leonard E.\ Parker Center for Gravitation, Cosmology, and Astrophysics, University of Wisconsin-Milwaukee, Milwaukee, WI 53201, USA}

	\author{Debnandini Mukherjee  \orcidlink{0000-0001-7335-9418}}
	\affiliation{NASA Marshall Space Flight Center, Huntsville, AL 35811, USA}
	\affiliation{Center for Space Plasma and Aeronomic Research, University of Alabama in Huntsville, Huntsville, AL 35899, USA}

	\author{Wanting Niu \orcidlink{0000-0003-1470-532X}}
	\affiliation{Department of Physics, The Pennsylvania State University, University Park, PA 16802, USA}
	\affiliation{Institute for Gravitation and the Cosmos, The Pennsylvania State University, University Park, PA 16802, USA}

	\author{Alex Pace}
	\affiliation{Department of Physics, The Pennsylvania State University, University Park, PA 16802, USA}
	\affiliation{Institute for Gravitation and the Cosmos, The Pennsylvania State University, University Park, PA 16802, USA}

	\author{Cort Posnansky}
	\affiliation{Department of Physics, The Pennsylvania State University, University Park, PA 16802, USA}
	\affiliation{Institute for Gravitation and the Cosmos, The Pennsylvania State University, University Park, PA 16802, USA}

	\author{Anarya Ray \orcidlink{0000-0002-7322-4748}}
	\affiliation{Leonard E.\ Parker Center for Gravitation, Cosmology, and Astrophysics, University of Wisconsin-Milwaukee, Milwaukee, WI 53201, USA}

	\author{Surabhi Sachdev \orcidlink{0000-0002-0525-2317}}
	\affiliation{School of Physics, Georgia Institute of Technology, Atlanta, GW 30332, USA}
	\affiliation{Leonard E.\ Parker Center for Gravitation, Cosmology, and Astrophysics, University of Wisconsin-Milwaukee, Milwaukee, WI 53201, USA}

	\author{Shio Sakon \orcidlink{0000-0002-5861-3024}}
	\affiliation{Department of Physics, The Pennsylvania State University, University Park, PA 16802, USA}
	\affiliation{Institute for Gravitation and the Cosmos, The Pennsylvania State University, University Park, PA 16802, USA}

	\author{Divya Singh \orcidlink{0000-0001-9675-4584}}
	\affiliation{Department of Physics, The Pennsylvania State University, University Park, PA 16802, USA}
	\affiliation{Institute for Gravitation and the Cosmos, The Pennsylvania State University, University Park, PA 16802, USA}

	\author{Ron Tapia}
	\affiliation{Department of Physics, The Pennsylvania State University, University Park, PA 16802, USA}
	\affiliation{Institute for Computational and Data Sciences, The Pennsylvania State University, University Park, PA 16802, USA}

	\author{Takuya Tsutsui \orcidlink{0000-0002-2909-0471}}
	\affiliation{RESCEU, The University of Tokyo, Tokyo, 113-0033, Japan}

	\author{Koh Ueno \orcidlink{0000-0003-3227-6055}}
	\affiliation{RESCEU, The University of Tokyo, Tokyo, 113-0033, Japan}

	\author{Aaron Viets \orcidlink{0000-0002-4241-1428}}
	\affiliation{Concordia University Wisconsin, Mequon, WI 53097, USA}

	\author{Leslie Wade}
	\affiliation{Department of Physics, Hayes Hall, Kenyon College, Gambier, Ohio 43022, USA}

	\author{Madeline Wade \orcidlink{0000-0002-5703-4469}}
	\affiliation{Department of Physics, Hayes Hall, Kenyon College, Gambier, Ohio 43022, USA}

	\maketitle

\section{Introduction}

According to the general relativistic description of a binary black hole (BBH), the two BH spins play an important role in determining the dynamics of the binary. Indeed, their magnitude and mutual interaction can speed up or slow down the gravitational wave (GW) emission and hence the rate of orbital shrinking. Moreover, if at least one of the two spins is mis-aligned with the orbital angular momentum~\cite{Apostolatos:1994mx}, the general relativistic interaction between the spin and the orbital angular momentum $\mathbf{L}$  causes the orbital plane to rotate around a constant axis. This phenomenon is called {\it precession} and the GW signal emitted in this case acquires a more complicated structure \cite{Kidder:1992fr, Kidder:1995zr, Buonanno:2002fy, Campanelli:2006fy}.

Despite the intricacies inherent in the modeling \cite{Racine:2008kj, Bohe:2012mr, Bohe:2015ana, Pratten:2020ceb, Estelles:2020osj, Akcay:2020qrj, Hamilton:2021pkf, Gamba:2021ydi, Ramos-Buades:2023ehm}, the underlying physical concept is clear: as GW emission attains its maximum along the direction $\hat{\mathbf{L}}$ of the orbital angular momentum, and given that $\hat{\mathbf{L}}$ evolves over time, an inertial observer will measure a time moduation of the wave amplitude and phase, as the orbital plane points towards and away from the observer.
This effect can be observed by current GW detectors.

Since the first detection of a GW from a BBH coalescence in 2015 \cite{LIGOScientific:2016vbw, LIGOScientific:2016vlm, LIGOScientific:2016gtq}, the LIGO~\cite{LIGOScientific:2014pky}, Virgo~\cite{VIRGO:2014yos} and KAGRA~\cite{KAGRA:2020tym} collaboration (LVK) have detected more than 90 GW signals from the first three observing runs~\cite{LIGOScientific:2018mvr, LIGOScientific:2020ibl, LIGOScientific:2021usb, KAGRA:2021vkt}.
The combined observations of these many signals established that a fraction of the existing binaries do {\it not} have the two spins aligned with the orbital angular momentum, although the degree of mis-alignment favors small values~\cite{LIGOScientific:2020kqk, KAGRA:2021duu}.
On the other hand, very few signals show strong imprints of precession, as quantified by the precessing inspiral spin parameter $\chi_P$  \cite{Schmidt:2014iyl}. Notable exceptions are GW190521 \cite{LIGOScientific:2020iuh}, GW191109 \cite{Zhang:2023fpp} and GW200129 \cite{Hannam:2021pit}, whose spin measurements may indicate support for large amount of spin-orbit mis-alignment. It is worth noting however that analyses on GW200129 are made difficult by data quality issues \cite{Payne:2022spz, Macas:2023wiw}, which could possibly undermine any statement about the precessing nature of the system.

The detection of precessing signals is of primary scientific interest, as the measurement of a strongly precessing signal provides an accurate measurement of the binary distance, breaking the degeneracy between distance and inclination, with direct impact on the accuracy of cosmological parameters inferred using GW observations \cite{Vitale:2018wlg, Yun:2023ygz}.
Moreover, precession measurements enable a better understanding of the spin distribution in the population of BBHs, which directly impacts our understanding of the BBH formation mechanism \cite{Farr:2017gtv, Farr:2017uvj, Johnson-McDaniel:2021rvv, Vitale:2022dpa}.
Indeed, BBH population studies indicate that heavily precessing systems are more likely to form through dynamic assembly rather than through common evolution \cite{Mapelli:2021taw}. However, currently the fraction of BBHs formed from each channel and a detailed understanding of these channels are still under debate and could possibly be illuminated by further detections of precessing signals.

The most recent LVK GW transient catalogues~\cite{LIGOScientific:2018mvr, LIGOScientific:2020ibl, LIGOScientific:2021usb, KAGRA:2021vkt} are built by performing sophisticated searches, which can be broadly categorized into unmodeled and modeled.
Unmodeled searches, such as those of the Coherent WaveBurst (cWB) pipeline \cite{Klimenko:2008fu,Klimenko:2011hz,Klimenko:2015ypf,Drago:2020kic}, search for transient signals with minimal assumptions by looking for coherent excess power between two detectors. Unmodeled searches tend to be less sensitive to compact binary coalescence (CBC) signals than modeled searches but can potentially identify signals from a broader range of sources. In this sense, unmodeled searches have a non-negligible sensitivity to precessing signals, even the most extreme ones.
On the other hand, modeled searches rely on matched-filtering \cite{Allen:2005fk} to correlate the detector's output with a large number of CBC waveform templates. While modeled searches are optimal when searching for known signals in purely Gaussian noise, they rapidly lose sensitivity for signals not included among the templates.

Despite having possibly detected a few precessing signals, the modeled searches deployed to build the modern catalogs \cite{LIGOScientific:2018mvr, LIGOScientific:2020ibl, LIGOScientific:2021usb, KAGRA:2021vkt, Nitz:2021zwj, Nitz:2021uxj, Olsen:2022pin, Mehta:2023zlk} search only for ``aligned-spin" systems and do not include the effects of precession in the models used for the template waveforms.
This choice drastically simplifies the searches, reduces the computational cost, and makes a systematic search of systems over a broad mass range feasible.
Clearly, aligned-spin searches do have some sensitivity to precessing systems, especially for mild precession, allowing for the statements about precession made for the BBH population~\cite{LIGOScientific:2020kqk, KAGRA:2021duu}.
Furthermore, given the low expected spin from binary neutron stars, the effects of precession are not expected to hurt GW search sensitivity significantly. Additionally, for the highest-mass systems (total mass ${\gtrsim 80}$ solar masses) detectable by the LVK, very few precession cycles of the waveform will be present so again, it is unlikely that a cumulative precession effect could significantly hurt detectability.
However, we expect precession effects to significantly impact the detectability of strongly precessing binary black hole mergers with a few 10s to 100s of cycles in the LIGO-Virgo senstivity band~\cite{Harry:2013tca, CalderonBustillo:2016rlt}.
For instance, as pointed out in~\cite{Dhurkunde:2022abc}, an aligned-spin search targeting neutron-star black-hole (NSBH) systems can miss up to $\sim 60\%$ of highly precessing sources with mass ratio $q\gtrsim 6$ and largely mis-aligned spins (i.e. $\chi_P> 0.5$).

Modeled searches for GWs are routinely performed by various matched-filter-based GW detection pipelines such as GstLAL \cite{Messick:2016aqy, Sachdev:2019vvd, Hanna:2019ezx, cannon2020gstlal, Ewing:2023qqe, Tsukada:2023edh}, PyCBC \cite{DalCanton:2014hxh, Usman:2015kfa, Nitz:2017svb, Davies:2020tsx}, the Multi-Band Template Analysis (MBTA) \cite{Adams:2015ulm, Aubin:2020goo}, the Summed Parallel Infinite Impulse Response (SPIIR) pipeline \cite{Luan:2011qx, Chu:2017ovg, Chu:2020pjv} among others \cite{Allen:2005fk, Privitera:2013xza,   Capano:2016dsf, Venumadhav:2019tad}.
Although they differ in their implementation details, they all rely on a multi-step algorithm which (i) correlates GW detector data with a large number of CBC waveform templates to find potential candidates (each called a {\it trigger}) and (ii) rank the triggers according to their false alarm rate (FAR), which determines how often we can expect such a trigger to occur by chance in a noise-only model.
Waveform templates are algorithmically placed in {\it template banks}, whose non-trivial generation has developed an active field of research \cite{Sathyaprakash:1991mt, Dhurandhar:1992mw, Owen:1998dk, Babak:2006ty, Cokelaer:2007mv, Harry:2009ea, Ajith:2012mn, Roy:2017oul, Coogan:2022qxs, Schmidt:2023gzj}.

The majority of these pipelines heavily rely on the simplicity of aligned-spin systems to obtain a convenient expression for the matched-filter and to reduce the number of templates required to cover the space of interest \cite{Owen:1998dk, Allen:2005fk}.
Indeed, while an aligned-spin system is characterized by 4 intrinsic parameters (i.e. two masses and the z-components of the two spins), a precessing system must be described by 8 intrinsic parameters (i.e. two masses and two 3D spin). For both systems, 7 additional extrinsic parameters are required to describe the position and orientation of the binary with respect to the observer.
For an aligned-spin system, the search statistics can be maximized over all 7 extrinsic parameters, reducing the template's dependency to only the 4 intrinsic parameters. However, for a precessing system (and/or a system where imprints from higher-order modes are considered), it is only possible to maximize the search statistics over a few extrinsic parameters. This leaves the template dependent not only on the 8 intrinsic parameters but also on the orbital inclination angle between the total angular momentum and the line of sight, and a reference phase~\cite{Harry:2016ijz, Harry:2017weg, Schmidt:2023gzj}.
Therefore, as the number of parameters required to describe a template increases from 4 to 10, the template-placement~\cite{Allen:2022lqr, Allen:2021yuy} becomes harder and delivers larger template banks, with an associated larger search computational cost.
This challenge, together with the need to introduce a modified expression for the matched filter, has limited the state-of-the-art pipelines to search only for aligned-spin systems.

Several approaches have been developed to tackle the challenges of precessing searches as listed above. Seminal works \cite{Apostolatos:1995pj, Apostolatos:1996rf, Buonanno:2002fy} focused on modeling precession with a set of phenomenological parameters, which were subsequently used to perform a matched-filtering search \cite{Buonanno:2005pt}. In this approach, the templates do not correspond to real physical signals but they are designed to match adequately physical precessing signals. More recently and in a similar fashion, \cite{McIsaac:2023ijd} leveraged the so-called two-harmonic approximation of a precessing signal \cite{Fairhurst:2019vut} to achieve a substantial sensitivity increase at a moderate computational cost when searching for precessing signals in the NSBH parameter space.

A different approach consists in directly using the physical waveform to filter the data \cite{DalCanton:2014qjd, Harry:2016ijz, Harry:2017weg}. This corresponds to a straightforward extension of the traditional matched-filter pipeline and has the benefits of using the real physical templates and, as such, it does not rely on any approximation of the waveforms, being applicable to virtually any type of binary system (including non-GR waveforms).
Employing this method typically requires a large number of templates and, as such, it poses the challenge of effectively generating large, high-dimensional template banks at a feasible computational cost.
In~\cite{Schmidt:2023gzj}, we addressed this challenge and developed a novel template placement method, specifically designed to generate large template banks, with minimal assumptions on the nature of the GW signal. Hence, by cutting the bank generation time, the exploration of the heavily precessing parameter space was made feasible.

Similar to the direct filtering method of~\cite{Harry:2017weg}, in this paper we design a search method for precessing signals using the features of the GstLAL search pipeline.
We demonstrate that our method results in an improvement in the pipeline's sensitivity to systems with large spin-orbit misalignment, reaching up to 100\% increase in the search sensitive volume, depending on the region of the parameter space.
We stress that with the new upgrades, the GstLAL pipeline is ready to handle not only a search for precessing signals but also a search for a generic BBH signals, such as those generated by eccentric binaries \cite{PhysRevD.102.043005, Ramos-Buades:2020eju} or those with imprints from higher-order modes (HOMs)~\cite{Blanchet:2008je, CalderonBustillo:2015lrt, Chandra:2022ixv}.

After briefly reviewing the physics of precessing binaries, in Sec.~\ref{sec:methods} we describe our search method and the modifications to the GstLAL pipeline needed to achieve our goal.
In Sec.~\ref{sec:banks}, after briefly reviewing the template bank generation method \cite{Schmidt:2023gzj}, we generate and validate two template banks, targeting heavily precessing BBHs.
We use these banks to conduct two searches using real data from the LIGO and Virgo interferometers during the third observing run (O3) \cite{LIGOScientific:2019lzm, KAGRA:2023pio}. In Sec.~\ref{sec:search_results}, we present the results from these searches, validate their performance and report the sensitivity improvement in comparison to the aligned-spin version of the pipeline.
In Sec.~\ref{sec:search_results}, we describe future work which will further improve the pipeline's sensitivity to precessing binary systems.
Final remarks are gathered in Sec.~\ref{sec:conclusion}.

\begin{figure*}[t]
	\centering
	\includegraphics[scale = 1.]{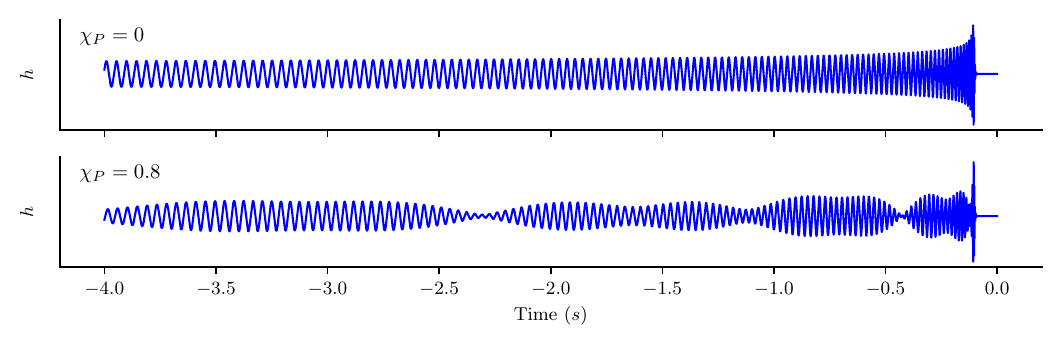}
	\caption{Effect of precession on a BBH signal. On the top panel we show the GW signal from an aligned-spin BBH, while in the bottom panel we show the effect of precession on the same system. On the $y$-axis we report the observed strain (with arbitrary units). The signal considered is characterized by $m_1 = \SI{40}{M_\odot}$ and $m_2 = \SI{3}{M_\odot}$ and $s_\text{1z} = -0.2$ and $s_\text{2z} = 0.7$. The precessing signal is chosen with $\chi_P = 0.8$, with precession placed only on the first BH spin. The system is in an ``edge on" configuration, with inclination ${\iota = \pi/2}$.}
	\label{fig:prec_WF_comparison}
\end{figure*}

\section{Search method} \label{sec:methods}

The rotation of the orbital plane induced by precession \cite{Apostolatos:1994mx} breaks the symmetry of the systems around the direction of the orbital angular momentum. This translates into the breaking of the symmetry between the $+$ and the $\times$ frequency-domain polarizations $\tilde{h}_+, \tilde{h}_\times$ of the GW generated by an aligned-spin systems which is typically expressed as
\begin{equation}\label{eq:symmetry_NP}
	\tilde{h}_+ \propto i \tilde{h}_\times.
\end{equation}
As most of the aligned-spin search pipelines use this symmetry to simplify the search problem~\cite{Allen:2005fk, DalCanton:2014hxh, Adams:2015ulm, Messick:2016aqy}, development of a precessing search pipeline must relax the assumption in Eq.~\eqref{eq:symmetry_NP} and tackle the general case.
In the following discussion, we describe the effects of precession on the GW signals and the consequences for the search method when we do not make any symmetry assumption about the two GW polarizations.

\subsection{Effects of precession on GW signals and searches} \label{sec:precession}

From a qualitative point of view, the effects of precession on a GW signal are more pronounced for heavily spinning asymmetric systems (large mass ratios) observed with an edge-on inclination (i.e. binaries observed with inclination angle $\iota \simeq \pi/2$) \cite{CalderonBustillo:2016rlt}.
As a binary system has suppressed GW emission on the $\iota \simeq \pi/2$ direction, even a small variation in the orbital plane orientation translates into a relatively large variation in GW emission.
Furthermore, precession is better observable in longer waveforms, primarily because there are more cycles during which precessing effects can accumulate. In typical data analysis applications, where the starting frequency of the signal is usually fixed, longer waveforms correspond to lower total mass. Consequently, with current instruments, the detection of precession is more likely in low-mass, highly asymmetric compact binary systems with very large spins \cite{Fairhurst:2019srr, Green:2020ptm}.

The features of the precessing waveforms are intimately tied to the number of templates needed to search for precession.
Indeed, as precession becomes more prominent, for instance at high mass ratios, the number of templates required to cover the parameter space increases significantly compared to aligned-spin searches.
Therefore, there is a trade-off between targeting regions where precession has the strongest impact on the signal and the computational cost associated with searching those regions with an adequate number of templates.
Similar considerations are crucial for selecting a region within the parameter space to target and it's essential to focus on areas where sensitivity improvement is more likely to be achieved while keeping a manageable computational cost.

Besides the two BH masses, $m_1, m_2$, a binary system is described by two 3D spin vectors $\mathbf{S}_1$ and $\mathbf{S}_2$. It is custom to introduce the {\it dimensionless} spin vectors, $\mathbf{s}_1$ and $\mathbf{s}_2$:
\begin{equation}
	\mathbf{s}_i = \frac{\mathbf{S}_i}{m_i^2}
\end{equation}
where each spin is normalized by the maximum allowed spin $S = m^2$ according to the Kerr solution of the Einstein equation \cite{Heinicke:2014ipp}.
Thus a generic precessing system is described by 8 {\it intrinsic} parameters, as opposed to an aligned-spin system, which is described by only 4 {\it intrinsic} parameters, i.e. the two masses and the two $\mathbf{L}$-aligned components of the spins, $s_\text{1z}$ and $s_\text{2z}$.
%



The effect of precession can be quantified by the {\it effective precession spin parameter} $\chi_P$ \cite{Schmidt:2012rh, Schmidt:2014iyl}, which can reduce the dimensionality of the parameter space.
Given a system with mass ratio $q = \frac{m_1}{m_2} \geq 1$ and spins $\mathbf{s}_1$ and $\mathbf{s}_2$, we define $s_{1\perp} = |\mathbf{s}_1 \times \hat{\mathbf{L}}|$ as the magnitude of the in-plane component of $\mathbf{s}_1$ and similarly for $s_{2\perp}$.
The effective precession spin parameter is equal to:
\begin{equation} \label{eq:chip}
	\chi_P = \max \bigg\lbrace s_{1\perp}, \frac{1}{q} \frac{4 + 3q }{ 3 + 4 q} s_{2\perp} \bigg\rbrace .
\end{equation}
$\chi_P$ defines a mapping between the original binary system (with two three-dimensional spins) and a system of reduced dimensionality, where ${\mathbf{s}_1 = (\chi_P, 0, s_{1z})}$ and ${\mathbf{s}_2 = (0,0,s_{2z})}$\footnote{Two such systems will show similar precessing effects, since by construction the average value of the precessing frequency is roughly the same.}.

$\chi_P$ is not always able to accurately capture the degree of precession and may prove inadequate for some heavily precessing systems or whenever HOMs are considered~\cite{Thomas:2020uqj}. However, despite the introduction of alternative precession measures in the literature \cite{Fairhurst:2019vut, Fairhurst:2019srr, Thomas:2020uqj, Gerosa:2020aiw}, $\chi_P$ retains its status as the most interpretable and widely utilized. Therefore, we will consistently employ $\chi_P$ throughout this paper to generate a template bank and to summarize our results.

The impact of spin misalignment can be amplified or suppressed by the inclination of the binary system, measured by the angle $\iota$ between the inclination of the observer's line of sight and the orbital angular momentum at a specific reference time. When $\iota$ is close to $0$ or $\pi$, there is minimal variation in the flux of emitted gravitational waves (GWs), resulting in a limited effect of precession on the waveform. Conversely, for $\iota \simeq \pi/2$, significant variations in the GW emission occur over time. Therefore, the inclination angle is a crucial parameter to consider when searching for precessing signals.

In Fig.~\ref{fig:prec_WF_comparison}, we show how precession changes the GW signal generated by a BBH. The upper panel reports an aligned-spin BBH, while the bottom panel shows the same signal with a non-zero $\chi_P$. We can see how precession changes the amplitude and phase evolution of the waveform. This work is concerned with signals such as the one depicted in the bottom panel of Fig.~\ref{fig:prec_WF_comparison}.


\subsection{Matched filtering}

The core of a matched-filtering pipeline relies on a frequentist test of hypotheses, to consistently accept or reject the hypothesis of a signal being in the data $d$ of a {\it single} interferometer.
More formally, we compare the ``signal" and the ``noise" hypotheses through the log likelihood ratio $\Lambda$ of the data under the two hypotheses:
\begin{equation}\label{eq:lambda}
	\Lambda = \log \frac{p(d\,|\,\text{signal})}{p(d\,|\,\text{noise})}.
\end{equation}

In the ``noise" hypothesis the data consist only of Gaussian noise $n(t)$, characterized by a known power spectral density (PSD) $S_n(f)$. In this case~\cite{Maggiore:2007ulw, Creighton_book}:
\begin{equation}\label{eq:noise_model}
	p(d\,|\,\text{noise}) \propto e^{\frac{1}{2}\rescalar{d}{d}}.
\end{equation}
In Eq.~\eqref{eq:noise_model}, $\rescalar{\cdot}{\cdot}$ denotes the {\it real} part of a complex scalar product ${\scalar{\cdot}{\cdot}}$ between two timeseries $a(t), b(t)$ defined as
\begin{equation}
	\scalar{a}{b} = 4 \int_{f_\text{min}}^{f_\text{max}} \!\!\!\! \d{f} \; \frac{\tilde{a}^*(f) \tilde{b}(f)}{S_n(f)}
\end{equation}
where the integral extends in a suitable frequency range $[f_\text{min}, f_\text{max}]$, $\tilde{\phantom{a}}$ denotes the Fourier transform and $*$ denotes complex conjugation.

In the ``signal" hypothesis, the data consists of a superposition of Gaussian noise $n(t)$ and a signal $h(t)$: ${d(t) = n(t)+ h(t)}$.
The probability $p(\text{s}|d)$ that a signal is present in some observed data, amounts to the test that the residual $d-h$ is consistent with Gaussian noise:
\begin{equation}\label{eq:signal_model}
	p(\text{s}|d) \propto e^{\frac{1}{2}\rescalar{d-h}{d-h}}.
\end{equation}
For a ground-based detector, the signal $h(t)$ depends on multiple physical parameters $\theta$, which characterize both the source ({\it intrinsic} parameters) and the position of the source with respect to the observer ({\it extrinsic} parameters):
\begin{equation}\label{eq:h_antenna_patterns}
	h(t; \theta) = F_+(\delta, \alpha, \Psi) h_+(t; \theta) + F_\times(\delta, \alpha, \Psi) h_\times(t;\theta).
\end{equation}
The functions $F_+, F_\times$, also called antenna pattern functions \cite{Finn:1992xs, Jaranowski:1998qm}, denote the interferometer response to the two polarizations of a GW $h_+, h_\times$, and they depend on the source's sky location given by right ascension $\alpha$ and declination $\delta$ and on the polarization angle $\Psi$. 
For a compact circular binary system, $h_+$ and $h_\times$ depend on the 8 intrinsic parameters (i.e. BH masses and spins), the inclination angle $\iota$, the reference phase $\varphi$, the luminosity distance of the source $D_L$ and the time of  coalescence $t_c$~\cite{Sathyaprakash_2009}.

Combining Eqs.~\eqref{eq:lambda}, ~\eqref{eq:noise_model} and~\eqref{eq:signal_model}, we find the likelihood ratio $\Lambda$ is given by:
\begin{equation}\label{eq:max_snr}
	\Lambda(\theta) = \rescalar{d}{\hat{h}}
\end{equation}
where $\hat{h} = \frac{h}{\rescalar{h}{h}^{1/2}}$.
%
The goal of a matched-filter search is then to maximize at every given time $\Lambda(\theta)$, also known as {\it search statistics}, as a function of both the intrinsic and extrinsic parameters $\theta$ of the model.
The maximisation problem $\max_{\theta} \Lambda(\theta)$ can be analytically solved for certain parameters, while a brute force approach is necessary for others.

\paragraph{The aligned-spin case}
As shown in Eq.~\eqref{eq:symmetry_NP}, in the case of aligned-spin circular orbit systems the two frequency-domain polarizations $\tilde{h}_+, \tilde{h}_\times$ are proportional to each other.
This can be used to show that the frequency-domain signal model can be written as \cite{Allen:2005fk, Harry:2016ijz}:
\begin{equation}\label{eq:simple_h_NP}
	\tilde{h}(f) = \frac{1}{D_\mathrm{eff}} e^{i\phi_0} \tilde{h}_+(f;m_1, m_2, s_\text{1z}, s_\text{1z})
\end{equation}
where $D_\mathrm{eff}$ and $\phi_0$ absorb the dependence of $h$ on the extrinsic parameters, including the inclination angle and reference phase.

As Eq.~\eqref{eq:max_snr} does not depend on an overall amplitude scaling of the template and we can analytically maximise over a constant phase shift, in the aligned-spin case, $\Lambda(\theta)$ only depends on four intrinsic parameters $m_1, m_2, s_\text{1z}, s_\text{1z}$, which we need to maximise over by brute force.
Maximizing Eq.~\eqref{eq:max_snr} over the phase $\phi_0$ in the template given by Eq.~\eqref{eq:simple_h_NP} gives:
\begin{equation}\label{eq:std_snr}
	\max \Lambda^2 = \lVert \scalar{s}{\hat{h}_+} \rVert^2 = \rescalar{s}{\hat{h}_+}^2 + \imscalar{s}{\hat{h}_+}^2
\end{equation}
where $\rescalar{\cdot}{\cdot}$ and $\imscalar{\cdot}{\cdot}$ denote the real and imaginary part of the scalar product, respectively.
The maximisation over the masses and spins is performed by evaluating Eq.~\eqref{eq:std_snr} for a discrete set of templates, gathered in a {\it template bank}. As discussed more in Sec.~\ref{sec:banks}, a template bank covers the parameter space of interest with waveform templates so that a random signal within that space will be recovered by at least one template with tolerable loss in the signal-to-noise ratio.
For each template in the bank, an aligned-spin matched-filtering pipeline evaluates Eq.~\eqref{eq:std_snr} as a function of time, by applying a time shift to the data $s$. The result of the operation is the so-called SNR timeseries $\mathrm{SNR}(t)$:
\begin{equation}
	\mathrm{SNR}(t) = |z(t)|.
\end{equation}
where $z(t)$ is the complex SNR timeseries
\begin{equation}
	z(t) = \rescalar{s}{\hat{h}_+}(t) + i \imscalar{s}{\hat{h}_+}(t) \label{eq:complex_snr}
\end{equation}
If we introduce $\hat{h}^{\pi/2}_+$ as $\pi/2$ phase shifted version of $\hat{h}_+$, such that in frequency domain $\tilde{h}_+ = i\tilde{h}^{\pi/2}_+$, the complex SNR timeseries becomes~\cite{Messick:2016aqy}:
\begin{equation}\label{eq:complex_snr_gstlal}
	z(t) = \rescalar{s}{\hat{h}_+}(t) + i \rescalar{s}{\hat{h}^{\pi/2}_+}(t)
\end{equation}
This is the expression used by the GstLAL pipeline  to filter the data.
In practice, this amounts to computing the correlation of the data with two {\it orthogonal} filters $\hat{h}_+$ and $\hat{h}^{\pi/2}_+$ and calculating the SNR by squaring and summing together the two contributions.

\paragraph{The precessing case}
In the case of a precessing signal, Eq.~\eqref{eq:simple_h_NP} is no longer valid since the two frequency-domain polarizations are not proportional to each other as they were in Eq.~\eqref{eq:symmetry_NP} for the aligned-spin case. In this case, we are still able to maximise over some of the extrinsic parameters but we lose the ability to analytically maximise over inclination $\iota$ and reference phase $\varphi$, which need to be included in the template bank.
As a consequence, a template bank for precessing signals covers a $10$ dimensional space, parametrized by the two BH masses, the two 3D spins as well as the angles $\iota$ and $\varphi$.

In this case, the analytical maximisation of Eq.~\eqref{eq:max_snr} yields \cite{Capano:2013raa, Schmidt:2014iyl, Harry:2017weg}:
\begin{equation}\label{eq:symphony_snr}
	\max \Lambda^2 = \frac{ \rescalar{s}{\hat{h}_+}^2 + \rescalar{s}{\hat{h}_\times}^2 -2\,\hat{h}_{+\times}\rescalar{s}{\hat{h}_\times}\rescalar{s}{\hat{h}_+}}{1- \hat{h}_{+\times}^2}
\end{equation}
where the parameter
\begin{equation}\label{eq:h_pc}
	\hat{h}_{+\times} = \rescalar{\hat{h}_+}{\hat{h}_\times}
\end{equation}
controls the discrepancy between the aligned-spin expression for the search statistic in Eq.~\eqref{eq:std_snr} and the general expression in Eq.~\eqref{eq:symphony_snr}.
Indeed, for $\hat{h}_{+\times} \to 0$, the latter equation reduces to the first, by using $\tilde{h}_+ \propto i \tilde{h}_\times$. Note that Eq.~\eqref{eq:symphony_snr} is fully generic and does not make any assumption about the nature of the signal. For this reason, besides precession, Eq.~\eqref{eq:symphony_snr} is also applicable to circular aligned-spin systems, where imprints of HOMs are considered.

A pipeline designed for detecting precessing and/or HOM signals must incorporate Eq.~\eqref{eq:symphony_snr} in the calculation of the SNR timeseries Eq.~\eqref{eq:complex_snr_gstlal}.
This update, combined with a precessing template bank and a suitable signal-consistency test, forms the key components for conducting a precessing search.
Neglecting to do so could lead to a substantial decrease in the recovered SNR, thus downgrading any potential discovery of a precessing signal and harming the pipeline sensitivity to precessing signals.

In this work, we compute the SNR timeseries for a precessing template as:
\begin{equation} \label{eq:complex_snr_symphony}
	z_\text{prec/HOM}(t) = \rescalar{s}{\hat{h}_+}(t) + i \rescalar{s}{\hat{h}_\perp}(t)
\end{equation}
where we introduced the ``orthogonalized" template\footnote{Note that, trivially, $\rescalar{\hat{h}_\perp}{\hat{h}_\perp} = 1$.}:
\begin{equation}\label{eq:h_orth}
	\hat{h}_\perp = \frac{1}{\sqrt{1- \hat{h}^2_{+\times}}} \left(\hat{h}_\times - \hat{h}_{+\times} \hat{h}_+\right)
\end{equation}
In App.~\ref{app:orthogonalization} we show that $|z_\text{prec/HOM}|^2$ is equivalent to the precessing/HOM search statistics Eq.~\eqref{eq:symphony_snr}.
Eq.~\eqref{eq:complex_snr_symphony} is particularly convenient as it enables the straightforward implementation of the search statistics in the GstLAL pipeline by simply modifying the filter waveforms.

It is worth noting that an alternative search statistics, suitable only for precessing signals, is introduced in~\cite{Harry:2016ijz}. While the alternative statistics has the obvious advantage of also maximising over the reference phase $\varphi$, thus reducing the dimensionality of the template bank. it comes at the price of a more complex functional form which translates into an increased computational cost of the search. Moreover, the statistics \cite{Harry:2016ijz} can be applied in a less general scenario as it does not apply to signals with HOM content.
Indeed by implementing the stastistics Eq.~\eqref{eq:complex_snr_symphony}, our pipeline is ready to search for signals with imprints from HOMs, thus reproducing the successful search presented in~\cite{Chandra:2022ixv}. See however~\cite{Wadekar:2023gea, Wadekar:2023kym, Wadekar:2024zdq} for a promising alternative method to search for systems with HOMs, which does not rely on Eq.~\eqref{eq:complex_snr_symphony}.

Whenever the value of search statistics passes a given threhshold (usually set to $4$), the time, the SNR value and the ringing template are recorded, together with some additional quantities: this is called a {\it trigger} and it constitutes the first step towards a detection.
Clearly, the search statistics itself cannot be used to claim any detection, as it only considers a single detector and may not be adequate in the presence of loud non-Gaussian noise features. Therefore, any detection claim cannot rely solely on selecting the loudest triggers but must instead rely on a more sophisticated ranking statistics~\cite{Nitz:2017svb, Hanna:2019ezx, Davies:2020tsx, Olsen:2022pin, Wadekar:2024zdq}, which takes into account other factors besides the measured values of the search statistics, possibly maximising or marginalising over the geometric parameters $\alpha$, $\delta$, and $\Psi$ in the multiple detector case (and not only in the single detector case).
As detailed below, the remaining steps of a search pipeline are devoted to build a suitable ranking statistics so as to assess whether a trigger has an astrophysical origin or it arised by random noise fluctuations.

\subsection{$\xi^2$ test}

The SNR is a very powerful statistics to rank triggers and discerning between GW signal and noise. Nevertheless, in the presence of loud transient noise (\textit{glitches}), the pipeline may record elevated values of SNR, thus producing false positive detections.
To mitigate this effect, the GstLAL pipeline conducts a signal consistency test \cite{Messick:2016aqy} for each trigger, denoted $\xi^2$. Note that it differs from the traditionally utilized $\chi^2$ signal consistency test \cite{Allen:2004gu}.
The value of $\xi^2$, together with the SNR, is used to identify candidate, leading to a more robust ranking.

The signal consistency test relies on the comparison between the measured complex SNR timeseries $z(t)$ and the expected timeseries $z_\text{pred} = z(0)R(t)$:
\begin{equation}\label{eq:chisq}
	\xi^2 = \frac{\int_{-\delta t}^{\delta t} \; \d{t} \;\; |z(t)-z(0)R(t)|^2}
		{\int_{-\delta t}^{\delta t} \; \d{t} \; \left( 2 - 2\left|R(t)\right|^2\right)}
\end{equation}
where the integrals extend on a short time window $[-\delta t, \delta t]$ around the trigger time and $R(t)$ is a suitable function describing the template response to a signal matching the template itself.

In the aligned-spin case, $R(t)$ is given by:
\begin{equation}\label{eq:R_std}
	R(t) = \rescalar{\hat{h}_+}{\hat{h}_+}(t) + i \rescalar{\hat{h}_+}{\hat{h}^{\pi/2}_+}(t)
\end{equation}
which can be easily obtained by plugging Eq.~\eqref{eq:simple_h_NP} into the complex timeseries Eq.~\eqref{eq:complex_snr_gstlal}.
%
%
In the precessing case, $R(t)$ has a more complicated expression \cite{Schmidt:2024kxy}:
\begin{align}\label{eq:R_prec_HOM}
	R_\text{prec/HOM}(t) =& \, \frac{1}{2} \left[ (\hat{h}_+|\hat{h}_+)(t) + (\hat{h}_\times|\hat{h}_\times)(t) \right] \nonumber \\
		&+ i \, \frac{1}{2}  \left[ (\hat{h}_+|\hat{h}_\times)(t) - (\hat{h}_\times|\hat{h}_+)(t) \right]
\end{align}
In Sec.~\ref{sec:chisq_results}, we discuss the performance of the adapted $\xi^2$ signal consistency test in a real search.

The expression for the template response $R_\text{prec/HOM}(t)$ is only approximate.
Indeed, as shown in \cite{Schmidt:2024kxy}, in the precessing/HOM case the true template response has a more complicated dependence on the two polarizations. This translates into an increased computational cost, and implementing the exact expression would require heavy changes to the pipeline. Luckily, \cite{Schmidt:2024kxy} demonstrates that for $SNR\lesssim 100$ the approximate template response Eq.~\eqref{eq:R_prec_HOM} is adequate for the purpose of the $\xi^2$ test, thus keeping the same simple simple factorization of Eq.~\eqref{eq:R_std}.
Therefore, since a signal has typically a much lower SNR, the use of the approximation is justified.

\subsection{Event identification stage} \label{sec:LR}

For each trigger, the GstLAL pipeline records the instrument $\{H1, L1, V1\}$, the trigger time $t$, the signal phase $\phi$ Eq.~\eqref{eq:simple_h_NP}, SNR, the $\xi^2$ value and the instantaneous sensitivity of each detector, measured in terms of horizon distance $D$.
Triggers happening on different instruments within a similar time window and with the same template are grouped together, forming a coincident trigger or {\it coincidence}.

As discussed above, to discriminate between coincidences arising from random fluctuations and coincidences of astrophysical origin, it is crucial to develop an accurate ranking statistics. 
Within the GstLAL framework, this is accomplished by computing the ratio between the probability that a coincidence is generated by a signal and the probability that a coincidence is generated by noise only~\cite{Cannon:2015gha, Hanna:2019ezx, Tsukada:2023edh}:
\begin{equation}\label{eq:LR}
	\mathcal{L} = \frac
			{p(\{\vec{D}, \vec{\rho}, \vec{\xi^2}, \vec{t}, \vec{\phi} \}\,|\, \text{signal})}
			{p(\{\vec{D}, \vec{\rho}, \vec{\xi^2}, \vec{t}, \vec{\phi} \}\,|\, \text{noise})}
\end{equation}
where $\rho$ is a shortand for SNR and $\{\vec{D}, \vec{\rho}, \vec{\xi^2}, \vec{t}, \vec{\phi}\}$ defines a coincidence of one or more instruments.
We call $p(\hdots \,|\, \text{signal})$ the {\it signal model}, while $p(\hdots \,|\, \text{noise})$ is called {\it noise model}.
The statistic Eq.~\eqref{eq:LR} is also known as likelihood ratio (LR).
Both the signal and noise model are parametric probabilistic models, whose parameters are set at the time of the search. They are conveniently factorized to exploit the known correlations between the recorded quantities --- see also \cite[Eqs.~(2) and (8)]{Tsukada:2023edh}.


The ranking statistics Eq.~\eqref{eq:LR} is used to assign a false alarm rate (FAR) to each coincidence. The FAR of a coincident event with LR $\bar{\mathcal{L}}$ corresponds to the expected rate of backgrounds events obtained with a LR $\mathcal{L} > \bar{\mathcal{L}}$. A detection is claimed whenever a candidate event reaches a given FAR threshold, usually set to $1/\text{year}$.

When it comes to precessing searches, the trigger time $t$, SNR, $\xi^2$ and $D$ keep the same physical meaning they have in the standard case and, for this reason, the terms in the LR involving only such quantities may not require any modification for the precessing case.
On the other hand, since Eq.~\eqref{eq:simple_h_NP} is not valid in the precessing case, the recorded phase $\phi$ looses its straightforward physical interpretation.
As a consequence, the factors of $\mathcal{L}$ depending on the phase may need to be adapted for a precessing search.
The phase dependent terms measure the coherence of the triggers among multiple detectors and in the GstLAL pipeline they correspond to $p(\vec{\rho}, \vec{t}, \vec{\phi} \,|\, \vec{D}, \text{signal})$ and $p(\vec{t}, \vec{\phi} \,|\, \vec{D}, \text{noise})$. We may refer to them as ``coherence" term.

Since we have no reason to believe that the ranking statistics, aside from the "coherence" term, is unsuitable for the precessing case, in this study we maintain without modification the ranking statistics utilized by the pipeline in its latest version~\cite{Tsukada:2023edh}.
However, the use of ``coherence'' terms designed for aligned-spin searches in the precessing case could potentially impact the search sensitivity negatively. Future work may investigate this impact on sensitivity and could develop an improved ranking statistic by computing and implementing an expression for the "coherence" term that is suitable for the precessing case.

\begin{table}
\footnotesize

	\begin{tabular}{c l l l l}
	\multirow{2}{*}{\phantom{Name}}  & \multirow{2}{*}{Mass range} &\multirow{2}{*}{Parameter space} &
	\multicolumn{2}{c}{Size} \\
	& & & AS & P \\
	 \toprule
 	 \vspace{0.05em} \\
	 Low $q \;\;$ &
	 	\begin{tabular}{@{}l@{}} $m_1, m_2 \in [8, 70] \mathrm{M_\odot}$ \\ $q\in [1,6]$  \\ \end{tabular}
	 		&
	 	\begin{tabular}{@{}l@{}}  $s_\text{1z} \in [-0.99, 0.99] $ \\ $s_\text{2z} \in [-0.99, 0.99] $ \\ $---------$ \\ $s_1 \in [0, 0.9]$ \\ $\theta_1 \in [-\pi, \pi]$ \\ $s_\text{2z} \in [-0.99, 0.99]$ \\ $\iota \in [0, \pi]$ \\ \end{tabular}
 	 		&
 	 	8425 & 1605625
	 	\\
	 \vspace{0.1em}  \\ \hline \vspace{0.1em} \\
	 High $q  \;\;$ &
	 	\begin{tabular}{@{}l@{}} $m_1\in [15, 70] \mathrm{M_\odot}$ \\ $m_2 \in [3, 10] \mathrm{M_\odot}$ \\ $q\in [5, 12]$ \\ \end{tabular}
	 		&
	 	\begin{tabular}{@{}l@{}} $s_\text{1z} \in [-0.99, 0.99] $ \\ $s_\text{2z} \in [-0.99, 0.99] $\\ $---------$ \\ $s_1 \in [0.5, 0.9]$ \\ $\theta_1 \in [-\pi, \pi]$ \\ $s_\text{2z} \in [-0.99, 0.99]$ \\ $\iota \in [0, \pi]$ \\ \end{tabular}
 	 		&
 	 	27016 & 2287083
	 	\\
	 \vspace{0.05em} \\
	 \bottomrule
	\end{tabular}
	\caption{We summarize here the most important features of the four template banks considered in this work. We select two regions of the parameter space, labeled ``Low $q$" and ``High $q$" respectively. For each region, we generate two template banks, one only gathering aligned-spin signals (AS) and another one including precessing signals (P). The aligned-spin banks covers the variable $M, q, \chi_\text{eff}$, while the precessing template bank cover the variables $M, q, s_1, \theta_1, s_\text{2z}, \iota$. For each template bank we report the range covered by each variable as well as the number of templates (size). The BBH variables not explicitly mentioned in the table are set to $0$.
	}
	\label{tab:template_bank}
\end{table}

\begin{figure*}[t]
	\centering
	\includegraphics[scale = 1.]{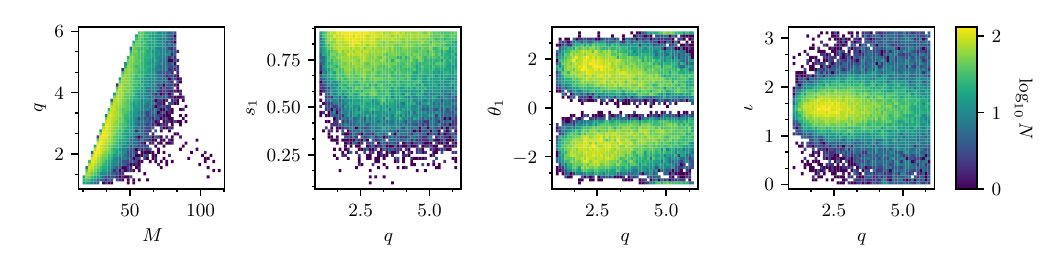}
	\caption{Binned templates of the ``Low $q$" {\it precessing} template bank. For each equal size bin, we color code the logarithmic number of templates so that the colour is also a measure of the template density. In the different panels, we consider the variables $M, q, s_1, \theta_1, \iota$. The bank was first introduced in \cite{Schmidt:2023gzj}.}
	\label{fig:bank_lowq_scatter}
\end{figure*}

\begin{figure*}[t]
	\centering
	\includegraphics[scale = 1.]{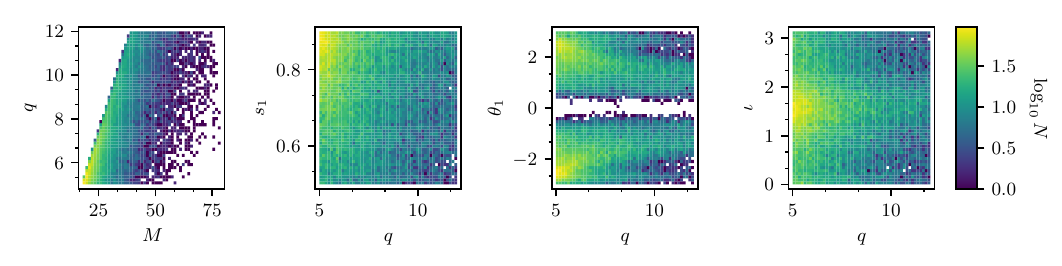}
	\caption{Binned templates of the ``High $q$" {\it precessing} template bank. For each equal size bin, we color code the logarithmic number of templates so that the colour is also a measure of the template density. In the different panels, we consider the variables $M, q, s_1, \theta_1, \iota$.}
	\label{fig:bank_highq_scatter}
\end{figure*}

\begin{figure*}[t]
	\centering
	\includegraphics[scale = 1.]{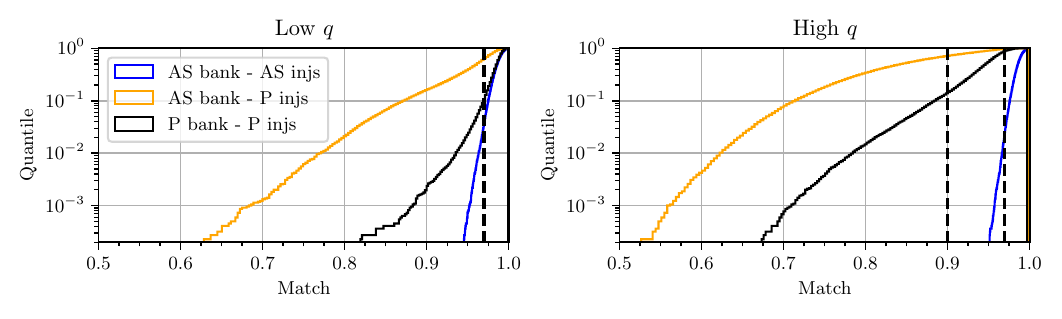}
	\caption{Fitting factor study for the four template banks considered in this work. For each given value of the match, we report the fraction of injections with fitting factor lower than than that value.
		The two aligned-spin banks (AS) and the two precessing banks (P) are tested against the relevant injection set of fully precessing injections used for searching the data. To evaluate their ability to cover the non-precessing space, the two AS template banks are also tested against a set of aligned-spin injections. The composition of the injection sets is described in the text.
		In the left panel, we report the results concerning the ``Low $q$" parameter space, while the right refers to the ``High $q$" parameter space. The vertical dashed lines mark the target minimal match of the template banks: $0.9$ for the ``High $q$" precessing bank and $0.97$ for the others.}
	\label{fig:banks_fitting_factor}
\end{figure*}

\begin{figure*}[t]
	\centering
	\includegraphics[scale = 1.]{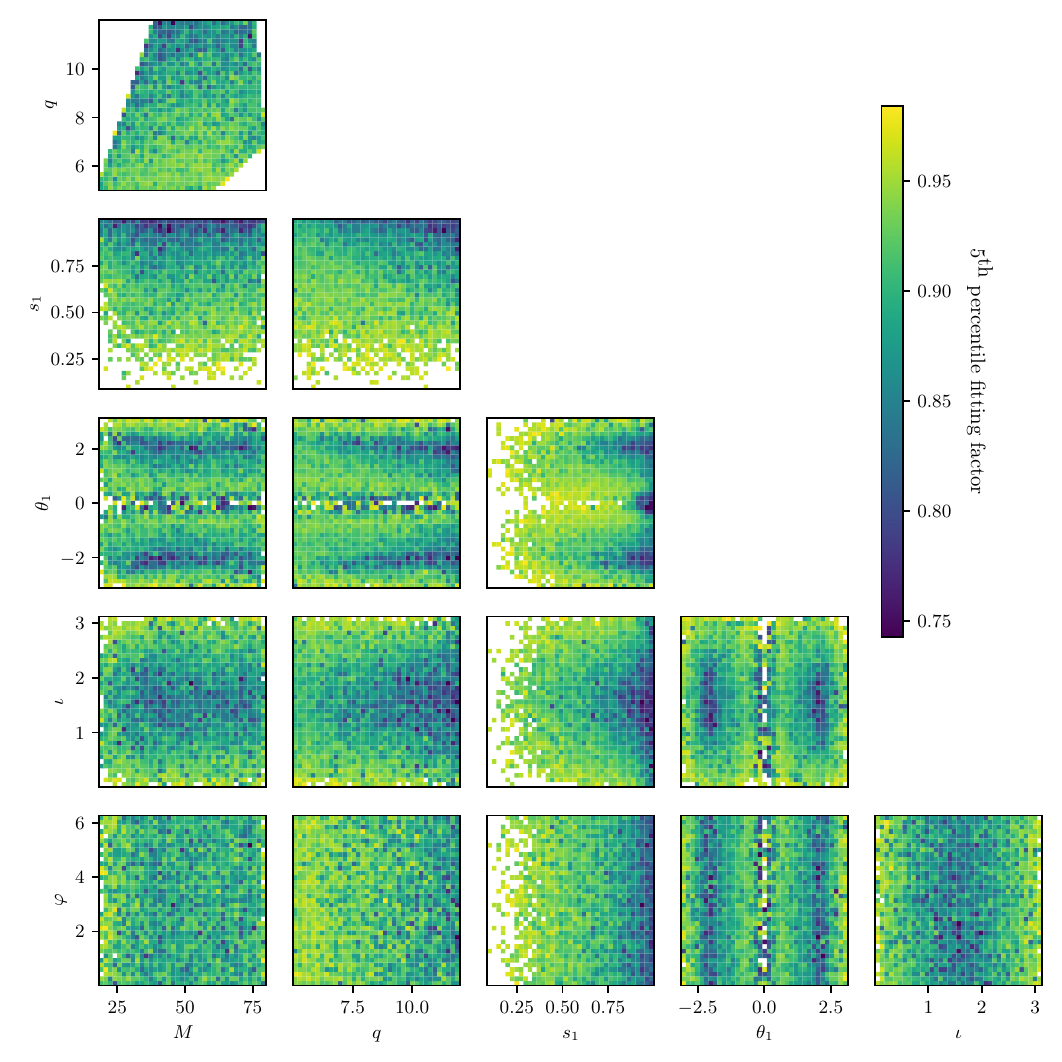}
	\caption{Fitting factor of the ``High $q$" precessing template bank, as a function of total mass $M$, mass ratio $q$, tilt $\theta_1$ and magnitude $s_1$ of the first BH, inclination angle $\iota$ and reference phase $\varphi$. For each bin, we report the $5^\text{th}$ percentile of the fitting factor distribution.}
	\label{fig:FF_high_q_precessing}
\end{figure*}

\section{Template banks} \label{sec:banks}

To demonstrate the efficacy of the updated GstLAL pipeline, we conduct two runs employing large template banks constructed for precessing signals. One is tailored for mildly asymmetric systems (referred to as the ``Low $q$" bank), while the other focuses on more asymmetric systems (the ``High $q$" bank).
For each precessing search, we run an aligned-spin search targeting the same mass range.
This comparative approach enables an assessment of the enhanced sensitivity resulting from the use of precessing templates and the modifications in the filtering scheme described in the previous section.
The next section Sec.~\ref{sec:search_results} presents the results of said comparison.
Here, we delve into the template banks' generation, providing insight into our choices for parameter space and validating their efficacy in covering the target regions. A summary of the template banks generated is presented in Tab.~\ref{tab:template_bank}.

For the purpose of template placement it is common to the define the {\it match} $|\scalar{\hat{h}_1}{\hat{h}_2}|$ between two normalized templates $\hat{h}_1, \hat{h}_2$, which serves as a measure of similarity between the two templates.
As it is standard in the literature, the size and performance of the bank are controlled by the {\it target minimal match} $MM$~\cite{Owen:1995tm}: a good bank should have only a small fraction $\lesssim 10\%$ of random signals with a match lower than $MM$ with the nearest template of the bank.
The actual fraction of signals with a match below $MM$ depends on the details of the template bank generation as well as on the parameter space of choice.

To generate the {\it aligned-spin} template banks, we use the state-of-the-art {\it stochastic} method \cite{Harry:2009ea, Ajith:2012mn}, which builds a template bank through an iterative process. At every step, a new template proposal is randomly sampled and the best match against the other templates of the bank is computed. The proposal is added to the bank only if its best match is lower than the target minimal match.
While the method delivers high quality template banks, it requires many expensive match computations. The number of match computation scales with the template bank size and with the dimensionality $D$ of the templates (i.e. the number of variables required to fully characterize a BBH template). If in the aligned-spin case the number of limited dimension ($D=4$) allows to generate a bank at a feasible cost, the same cannot be said for a precessing template bank, where the larger number of dimensions make the stochastic method too costly.

To avoid these shortcomings, we generate the {\it precessing} template banks with our publicly available code \texttt{mbank} \cite{Schmidt:2023gzj, mbank}, specifically designed to cover high dimensional BBH parameter spaces, such as those associated to precession.
To place templates, it employs a {\it metric} approximation to the distance between templates, which is used to meaningfully define a volume element on the parameter space: the number of templates placed by the algorithm in a given volume $\mathcal{T}$ is proportional to the volume of $\mathcal{T}$, computed by the metric.
\texttt{mbank} implements a novel metric expression, specifically designed to provide an approximation to Eq.~\eqref{eq:symphony_snr}, rather than to Eq.~\eqref{eq:std_snr} as done by other metric codes \cite{owen_metric, Messenger:2008ta, Prix:2007ks, Brown:2012qf, Coogan:2022qxs, Hanna:2022zpk}.
Templates are added to the template banks \cite{Messenger:2008ta, Coogan:2022qxs} by sampling from a suitably trained normalizing flow model \cite{norm_flow, nflows_paper, Kobyzev_2021}, without any control on their mutual distance. The bank's size grows until a certain fraction $\eta < 1$ of the parameter space is covered by templates.
Our placing method, known in the literature as {\it random}, ensures very fast template placement at the price of up to $\sim 50\%$ larger banks than those produced by the state-of-the-art stochastic approach. While this may seem sub-optimal, several studies \cite{Messenger:2008ta, Allen:2021yuy, Allen:2022lqr} argued that in the limit of a large number of dimension, the random method reaches optimality.

As it is common, the non precessing template banks sample the variables
\begin{equation*}
	M, q, s_\text{1z}, s_\text{2z}
\end{equation*}
while both the precessing templates bank sample the variables
\begin{equation*}
	M, q, s_1, \theta_1, s_{2z}, \iota
\end{equation*}
where $M = m_1 + m_2$ denotes the total mass of the binary and the first spin is expressed in spherical coordinates:
\begin{align}
	s_\text{x} & = s \sin\theta \\
	s_\text{y} & = 0 \\
	s_\text{z} & = s \cos\theta.
\end{align}
In our template bank, precession is only encoded into $s_\text{x}$\footnote{Here $s_\text{x}$ plays the role of the precession spin parameter $\chi_P$.}, while all the other in-plane spin components $s_\text{1y}, s_\text{2x}, s_\text{2y}$ are set to zero.

Note that we set the reference phase $\varphi = 0$, despite the fact that the search statistics Eq.~\eqref{eq:symphony_snr} does not explicitly maximises over $\varphi$.
The motivation for this choice is mainly technical: the metric approximation used by \texttt{mbank} is nearly degenerate if $\varphi$ is considered, thereby limiting the algorithm's ability to effectively cover the space and resulting in a bank with more templates than needed. This issue was also discussed in~\cite{Schmidt:2023gzj}.
Moreover, we find that omitting the reference phase $\varphi$ does not significantly impact the bank's ability to recover a generic signal.
This is demonstrated by a performance study presented in Fig.\ref{fig:FF_high_q_precessing}, where the fitting factor does not depend on the reference phase $\varphi$, and in Fig.~\ref{fig:banks_fitting_factor}, where the performance of the precessing banks, assessed with injections with a generic reference phase $\varphi$, is consistent with that of the aligned-spin banks. This indicates that our choice to neglect $\varphi$ has minimal effect on the effectualness of the template bank.

To generate the precessing template banks, we use the frequency domain approximant \texttt{IMRPhenomXP} \cite{Pratten:2020ceb}, while for the aligned-spin banks we use \texttt{IMRPhenomD} \cite{Khan:2015jqa}. In all cases, we consider frequencies between ${f_\text{min} = \SI{15}{Hz}}$ and ${f_\text{max} = \SI{1024}{Hz}}$ and we use the Advanced LIGO O4 Design Power Spectral Density (with $\SI{190}{Mpc}$ range) \cite{O4_PSDs}.

To validate a template bank, we select a number of test signals (also called injections) and for each signal $s(\theta)$, characterized by $\theta$, we compute the {\it fitting factor} $FF$:
\begin{equation}\label{eq:FF}
	FF(\theta) = \max_{\theta^\prime \in \text{bank}} \mathcal{M}(\theta, \theta^\prime)
\end{equation}
where the {\it match} $\mathcal{M}$ is computed with Eq.~\eqref{eq:symphony_snr} evaluated for a normalized signal $\hat{s}(\theta)$ and a normalized template $\hat{h}_+(\theta^\prime)$, $\hat{h}_\times(\theta^\prime)$.

The precessing injections are sampled from the relevant mass space using a probability distribution uniform in $m_1$ and $q$. The spins' magnitudes are sampled so that the fourth power of spin is uniformly distributed in $[0,1]$. While this distribution of spins is clearly unphysical, it was chosen to make sure that a large fraction of the injected singals had a large precessing spin: in this way, approximately $70\%$ of the injected signals have $\chi_P>0.5$.
The spin directions, the sky location and the binary orientation are all isotropically distributed over the solid angle. The reference phase is uniformly sampled in $[0, 2\pi]$.
We generate a set of $21947$ precessing injections, with a $30$ seconds spacing. This will be used both for validating the banks and for the runs described in Sec.~\ref{sec:search_results}.
To validate the aligned-spin banks, we use the {\it same} $21947$ precessing injections as before, with the in-plane spin components set to zero (i.e. ${s_{1x} = s_{1y} = s_{2x} = s_{2y} = 0}$).

\begin{figure*}[t]
	\centering
	\includegraphics[scale = 1.]{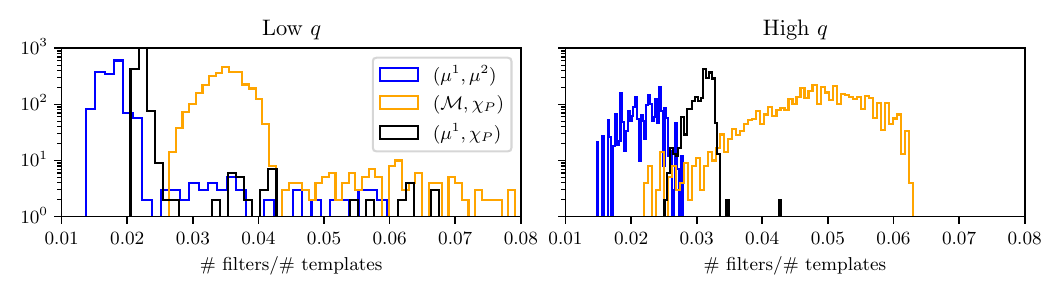}
	\caption{Compression achieved by the Singular Value Decomposition of the templates of the two precessing template banks considered in this work. For each bin, the compression is measured by the ratio between the number of filters and the number of templates. The number of filters is obtained by averaging, across different time slices, the number of SVD basis. The average is weighted by the sampling rate of each time slice. For different choices of the sorting variables $(\alpha^{1}, \alpha^{2})$ used to construct the bank split, we report the histogram of the compression achieved for different bins.
		}
	\label{fig:svd_compression}
\end{figure*}

\paragraph{A template bank for low mass ratio systems} \label{par:low_q_bank}

The precessing ``Low $q$" template bank was introduced in \cite[Sec. V]{Schmidt:2023gzj}. It is a large bank covering systems with component masses $m_1, m_2 \in [8, 70] \mathrm{M_\odot}$ with mass ratio limited into $q\in [1,6]$.
With a target minimal match of $0.97$, the template bank has $1.6$ million templates.
We also generated an aligned-spin bank in the same mass region, consisting of $1.4 \times 10^4$ templates. A summary of the two ``Low $q$" banks is reported in Tab.~\ref{tab:template_bank}.

In Fig.~\ref{fig:bank_lowq_scatter} we plot the template density of the precessing bank, while in Fig.~\ref{fig:banks_fitting_factor} we present the results of an injection study.
We note that both the precessing and the aligned-spin banks have an acceptable performance with only $\sim 10\%$ of the test signals having a match below the $0.97$ threshold.
The aligned-spin bank recovers $\sim 15\%$ of the precessing injections with a fitting factor below $0.9$. This suggests that the aligned-spin bank approximates a $0.9$ target minimal match bank for precessing signal. This remark will be needed to explain some of the search results.

As discussed in \cite{Schmidt:2023gzj}, the precessing bank provides a very poor coverage in the ``low $q$, low $M$", characterized by $q \lesssim 1.2$ and $M \lesssim \SI{20}{\mathrm{M_\odot}}$. For this reason, we exclude such region from our searches (and also from our injection studies)\footnote{To address the issue of poor coverage, an effective strategy could involve using aligned-spin templates to cover the ``low $q$, low $M$'' region.}: systems in this region are expected to show little or no effect from precession, hence this region is of scarce interest for our purpose of quantifying the benefit of using precessing templates.
Finally, we note in Fig.~\ref{fig:bank_lowq_scatter} that two small regions at $\theta = \pm \pi$ and $q \simeq 5$, corresponding to the aligned-spin limit, are unphysically populated by many templates. This is a feature of the metric approximation, probably due to the approximant in use, and was also discussed in \cite{Schmidt:2023gzj}.


\paragraph{A template bank for high mass ratio systems} \label{par:high_q_bank}

To target heavily asymmetric precessing systems, we construct the precessing ``High $q$" template bank. The bank targets systems with mass ratio $q \in [5, 12]$, with primary mass $m_1\in [15, 70] \mathrm{M_\odot}$ and secondary mass $m_2 \in [3, 10] \mathrm{M_\odot}$.
For the precessing bank, we limit the magnitude of the primary spin $s_1$ to the range $[0.5, 0.9]$. As shown in the fitting factor study Fig.~\ref{fig:banks_fitting_factor}, our choice does not impact the bank injection recovery, even in the low $s_1$ region excluded by the bank.
As part of our development, we also generated a bank covering the entire spin range. However, this additional bank, consisting of $3\times 10^5$ more templates, did not demonstrate improved coverage of the low $s_1$ region.

Given the vast size of the parameter space, employing a template bank with a target minimal match of $0.97$ would entail dealing with an unwieldy number of templates, reaching as high as $30$ million\footnote{This number comes from an attempt we made to generate such bank with \texttt{mbank}. A similar number appears in \cite{McIsaac:2023ijd}, although referring to a template bank covering a slightly lower mass range.}. As modern pipelines are unable to handle banks of similar size, we have pragmatically set a target minimal match requirement of $0.9$ for the precessing bank.
With the bank now comprising $2.3$ million templates, this adjustment helps manage the computational load while still ensuring a reasonable level of template coverage in the search for GW signals.

We also generate an aligned-spin bank, covering the same mass range: with a target minimal match of $0.97$, the aligned-spin bank has $3.9 \times 10^4$ templates.
 It's noteworthy that the aligned-spin bank is two orders of magnitude smaller than the precessing bank. Consequently in this case, a precessing search is nearly two orders of magnitude more computationally demanding than its aligned-spin counterpart.
Other details of the two ``High $q$" are reported in Tab.~\ref{tab:template_bank}.
Fig.~\ref{fig:bank_highq_scatter} shows the template density of the precessing bank, where we observe that the bulk of the templates are placed in the high $q$, high $s_1$ region, particularly for systems with $\theta_1 \simeq \pm \pi/2$, as seen with an inclination $\iota \simeq \pi/2$. This distribution is consistent with the fact that signals in such regions exhibit the largest precession content.
In Fig.~\ref{fig:banks_fitting_factor}, we report the results of an injection study for the two banks, showing that both the precessing and ``aligned-spin'' bank are able to satisfy the relevant minimal match requirements of $0.9$ and $0.97$ respectively.

In Fig.~\ref{fig:FF_high_q_precessing}, we study the dependence of the fitting factor of the precessing template bank on the several parameters characterizing a template. We note that the fitting factor is pretty stable across different regions of the parameter space, confirming the quality of our template bank.
As the template bank only covers values of $s_1$ up to $0.9$, we expect and observe a drop in accuracy at large values of the primary spin, which lie outside the template bank. The reduced accuracy at high spins seems to be limited only to edge-on systems (i.e. $\iota \simeq \pi/2$) with large values of $\chi_P$ (i.e. $\theta \simeq \pm\pi/2$).
Finally, we see that the coverage is rather uniform in $\varphi$. This justifies our choice of not including the reference phase as part of the variables characterizing the precessing template banks.

For both the precessing and the aligned-spin bank, only $\sim 10 \%$ of the injections are below the target minimal match threshold of $0.9$ and $0.97$ respectively. This is consistent with the fitting factor results obtained in the ``Low $q$" case.
With only $\sim 20\%$ of the precessing injections recovered with a match higher than $0.9$, the aligned-spin bank struggles to cover the precessing parameter space satisfactorily.
Note that in the ``Low $q$" case, the aligned-spin template bank was able to recover $\sim 85\%$ of the precessing injections with a match higher than $0.9$.

\subsection{SVD compression} \label{sec:svd_compression}

As opposed to the large majority of other pipelines, the GstLAL pipeline performs the matched filtering in time domain. As computing the cross-correlation between two timeseries in time domain is notoriously more expensive than in frequency domain\footnote{Indeed, for a timeseries of $D$ points, computing the correlation in time domain has a computational cost of $\mathcal{O}(D^2)$ while in frequency domain the same operation amounts to computing a Fast Fourier transform and has a cost of $\mathcal{O}(D\log D)$.}, a Singular Value Decomposition (SVD) based template bank compression \cite{Cannon:2010qh, Cannon:2011vi, Messick:2016aqy} scheme has been implemented to mitigate the cost of filtering.

Templates in the banks are grouped together in bins of $\mathcal{O}(500)$ templates. Each template is then divided in different time slices, each with a different sample rate. For each time slice of each bin an SVD decomposition of the templates is performed. The data are then filtered only using a small subset of the SVD basis and the SNR timeseries for each template is reconstructed by ``inverting" the SVD.
Filtering the data using the SVD basis results in a reduction of one or two orders of magnitude of the number of filters.
Clearly, the bins for the SVD must be carefully chosen: heuristically, each bin should gather very similar waveforms, so that only a small number of SVD basis are required to faithfully represent the templates.

The bins are constructed based on two quantities $\alpha^{1}, \alpha^{2}$.
As a first step, the template bank is split into $N_\text{groups}$ bins based on the $\alpha^{2}$ values. Second, groups of $N_\text{tmplt}$ templates are gathered together based on their $\alpha^{1}$ value to form an SVD bin.
Several choices have been implemented for the two sorting variables $(\alpha^{1}, \alpha^{2})$. Common choices for $\alpha^{1}$ are the template duration or the chirp mass $\mathcal{M}$:
\begin{equation}\label{eq:mchirp}
	\mathcal{M} = \frac{(m_1 m_2)^{3/5}}{(m_1 + m_2)^{1/5}}
\end{equation}
A typical choice for $\alpha^{2}$ is the effective spin parameter $\chi_\text{eff} = \frac{m_1 s_\text{1z} + m_2 s_\text{2z}}{m_1 + m_2}$ \cite{Maggiore:2007ulw}. As in the case of precessing templates, $\chi_\text{eff}$ does not capture the effect of precession, the effective precession spin parameter $\chi_P$ can be a good option for $\alpha^{2}$ instead of $\chi_\text{eff}$.
In \cite{Sakon:2022ibh}, the authors suggests to use the PN variables $\mu^1$ and $\mu^2$, first introduced in \cite{Morisaki:2020oqk}, and they achieve very high compression.

In Fig.~\ref{fig:svd_compression} we investigate how the compression provided by the SVD varies for the two template banks and for different choices of sorting quantities. The study aims at finding the best sorting quantities for maximal speed up.
In our study, we set $N_\text{tmplt}= 500$ and $N_\text{groups} = 20$.

Our results show that the sorting with variables $\mu^1$ and $\mu^2$ introduced in \cite{Sakon:2022ibh} provides the best speed up even in the precessing case.
Regardless of the sorting variables considered, the ``Low $q$" precessing bank shows better compression than the ``High $q$" precessing bank. This is expected on the basis that the ``High $q$" bank gathers signal with higher precession content and more complex morphology. Therefore, a higher number of SVD basis is needed to faithfully reproduce the waveform.

A comparison with the results in \cite{Sakon:2022ibh} shows that, in the precessing case, the SVD compression is a factor of $2/3$ worse than the align spin case. As noted before, this is expected due to the increased complexity of precessing waveforms. The higher number of SVD basis translates directly into a larger computational cost and it needs to be taken into account when allocating the computational resources for a precessing search.

\begin{figure}[t]
	\centering
	\includegraphics[scale = 1.]{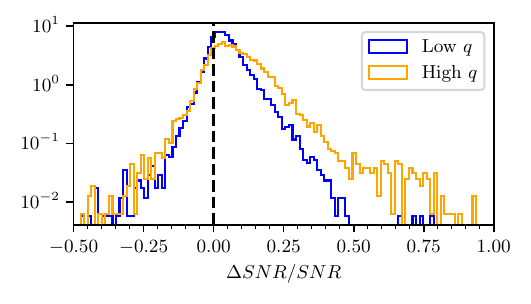}
	\caption{Histogram with the fractional SNR gain of a precessing search over its aligned-spin counterpart. For the two cases ``Low $q$" and ``High $q$", we report the discrepancy $\Delta \text{SNR} = \text{SNR}_\text{precessing}-\text{SNR}_\text{aligned-spin}$ of the SNR recovered by the precessing and the aligned-spin, normalized by the SNR of each injection. The injections are described in Sec.~\ref{sec:banks} and the data refer to the LIGO Livingston detector.
		}
	\label{fig:hist_snr_diff}
\end{figure}

\section{Precessing Searches Results} \label{sec:search_results}

Using the four template banks described in the previous Sec.~\ref{sec:banks}, we run four searches on the publicly available LIGO and Virgo data~\cite{KAGRA:2023pio} obtained through GWOSC~\cite{Vallisneri:2014vxa}, taken during the third observing run O3 between GPS times $1259423400$ and $1260081799$.
As described above, we generate two sets of precessing injections, one for the ``Low $q$" banks and another for the ``High $q$" banks. This choice allows us to directly compare the results of a precessing search with those of an aligned-spin search.


When assessing the sensitivity of a precessing search compared to an aligned-spin search, two competing effects come into play. On one hand, the precessing search enhances the recovered SNR and lowers the $\xi^2$, leading to improved sensitivity. On the other hand, the larger bank's size results in an increased number of background triggers (i.e., false alarms), which can potentially downgrade the significance of any candidate event.
Determining whether the improvement in recovered SNR outweighs the increased background triggers, and thus whether a precessing search indeed provides enhanced sensitivity will be the focal point of this section.

In what follows, we evaluate the improvement brought by a precessing search using three different figure of merits: recovered SNR, results from the $\xi^2$ test and, finally, increase in search sensitivity.
To study the sensitivity we used the same injection set introduced for bank validation in Sec.~\ref{sec:banks}, with injections sampled uniformly in $m_1$ and mass ratio inside the mass ranges covered by each template bank.

\begin{figure}[t]
	\centering
	\includegraphics[scale = 1.]{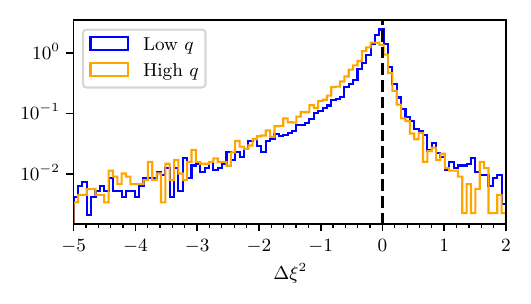}
	\caption{Histogram with the difference between the $\xi^2$ measured by a precessing search and its aligned-spin counterpart. For the two cases ``Low $q$" and ``High $q$", we report the discrepancy $\Delta \xi^2 = \xi^2_\text{precessing}-\xi^2_\text{aligned-spin}$, evaluated on the set of precessing injections described in Sec.~\ref{sec:banks}. Data refer to the LIGO Livingston detector.
		}
	\label{fig:hist_chisq_diff}
\end{figure}

\subsection{Recovered SNR} \label{sec:snr_results}

In Fig.~\ref{fig:hist_snr_diff}, we present a histogram for each injection, illustrating the discrepancy $\Delta SNR$ between the SNR recovered by the precessing search and the SNR recovered by the aligned-spin search. The SNR difference is divided by the nominal SNR of the injection and we the data refers both to the ``Low $q$" and ``High $q$" cases.
This analysis offers a valuable metric to evaluate the improvement brought by filtering the data with precessing templates compared to aligned-spin ones.

A certain degree of scattering of the discrepancies in recovered SNR ($\Delta SNR$) is expected due to random noise fluctuations. However, our analysis reveals a systematic trend: searches utilizing precessing templates consistently recover a larger fraction of the injected SNR compared to those using aligned-spin templates. This improvement can be substantial, with precessing searches recovering up to $75\%$ more SNR than their aligned-spin counterparts.
The systematic increase in recovered SNR fraction with precessing templates aligns with expectations, given the closer match of precessing templates to injected waveforms shown in Fig.~\ref{fig:banks_fitting_factor}.
Whether the improved SNR recovery translates into an larger search sensitivity will be assessed in the next sections.

Finally, we observe that the improvement brought by the ``High $q$" precessing search over its aligned-spin counterpart is larger than in the ``Low $q$" case. This is consistent with our results in Fig.~\ref{fig:banks_fitting_factor}, which show that the aligned-spin ``High $q$" template bank exhibits poorer performance than the ``Low $q$" one, in recovering precessing injections. This is a direct consequence of the effects of precession being more prominent for asymmetric systems.


\begin{figure*}[t]
	\centering
	\includegraphics[scale = 1.]{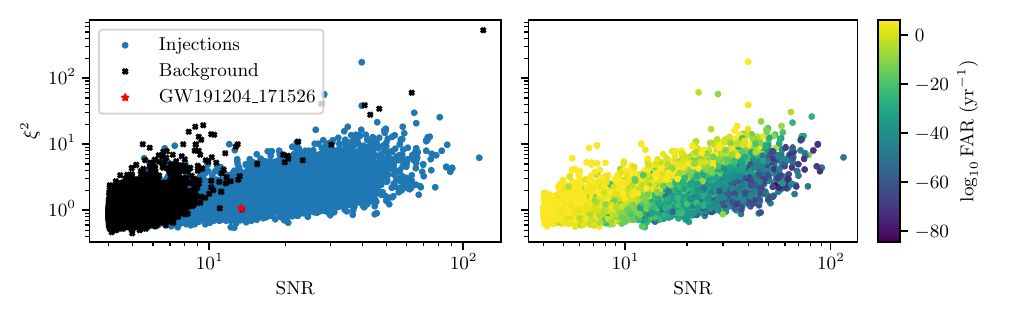}
	\caption{``SNR-$\xi^2$" plot for the ``High $q$" aligned-spin search. On the left panel, we report the SNR and $\xi^2$ of the background triggers, not associated to any injection, and the injection triggers, corresponding to an injection. On the right panel, we only report the injection triggers, colored by the logarithm of the FAR assigned by the pipeline. Data refer to the LIGO Livingston detector. The red star refers to the trigger produced by the GW event GW191204\_171526 \cite{KAGRA:2021vkt}, detected by our search with $FAR < \frac{1}{1000} \SI{}{yr^{-1}}$.}
	\label{fig:snr_chisq_plot_NP}
\end{figure*}

\begin{figure*}[t]
	\centering
	\includegraphics[scale = 1.]{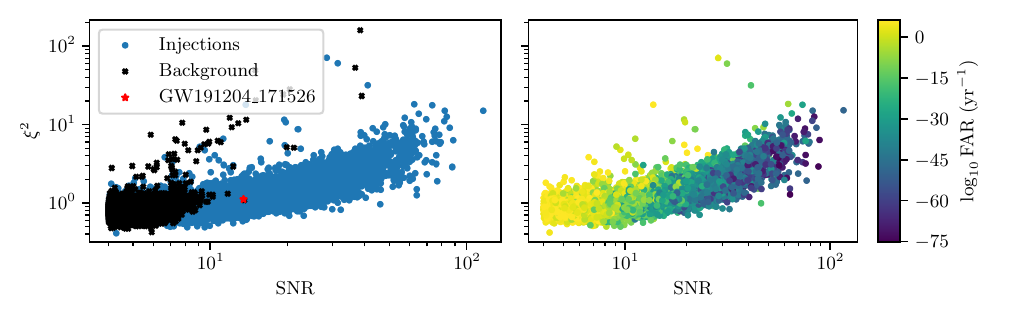}
	\caption{``SNR-$\xi^2$" plot the ``High $q$" precessing search. On the left panel, we report the SNR and $\xi^2$ of the background triggers, not associated to any injection, and the injection triggers, corresponding to an injection. On the right panel, we only report the injection triggers, colored by the logarithm of the FAR assigned by the pipeline. Data refer to the LIGO Livingston detector. The red star refers to the trigger produced by the GW event GW191204\_171526 \cite{KAGRA:2021vkt}, detected by our search with $FAR < \frac{1}{1000} \SI{}{yr^{-1}}$.}
	\label{fig:snr_chisq_plot_P}
\end{figure*}





\begin{figure*}[t]
	\centering
	\includegraphics[scale = 1.]{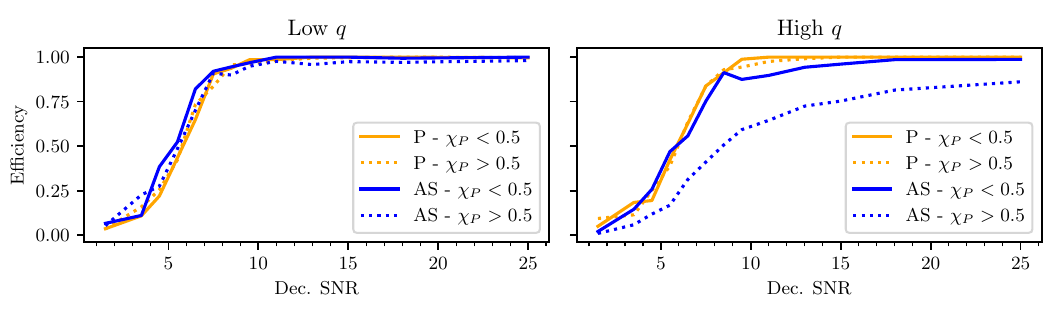}
	\caption{Efficiency of the BBH searches as a function of the loudest SNR measured by the three detectors, also called the ``Decisive SNR". The efficiency is defined as the fraction of signal detected at constant $FAR = 1/10\text{ years}$. In each panel, we show the efficiency pertaining the aligned-spin (AS) and the precessing (P) searches, reported separately for $\chi_P>0.5$ and $\chi_P<0.5$. The left panel refers to the ``Low $q$" search, while the right panel to the ``High $q$" search.}
	\label{fig:efficiency_vs_SNR}
\end{figure*}

\begin{figure*}[t]
	\centering
	\includegraphics[scale = 1.]{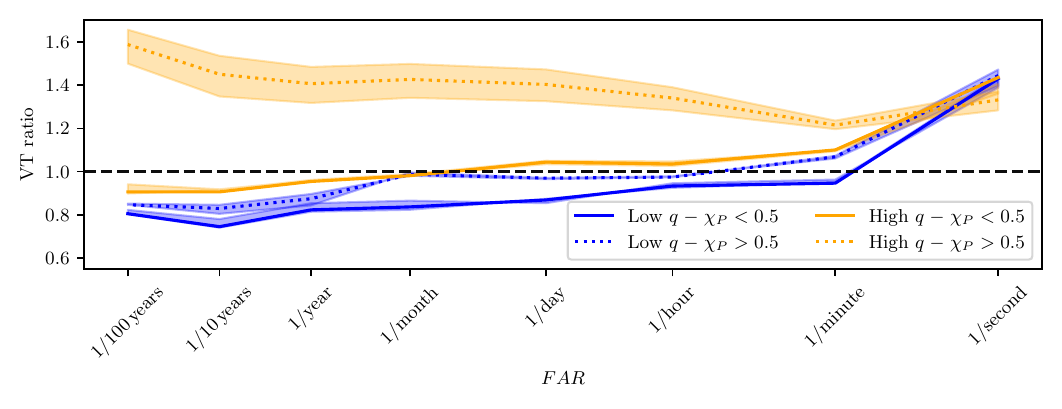}
	\caption{Sensitive volume increase as a function of the FAR detection threshold. The increase in sensitive volume is computed in terms of the VT ratio, measured between the VT of a precessing search and the VT of an aligned-spin searches. The two blue lines refer to the ``Low $q$" region, while the blue lines correspond to ``High $q$". A dashed line refers to the VT computed only from injections with $\chi_P>0.5$, while a solid line corresponds to $\chi_P<0.5$.}
	\label{fig:vt_vs_far}
\end{figure*}

\begin{figure*}[t]
	\centering
	\includegraphics[scale = 1.]{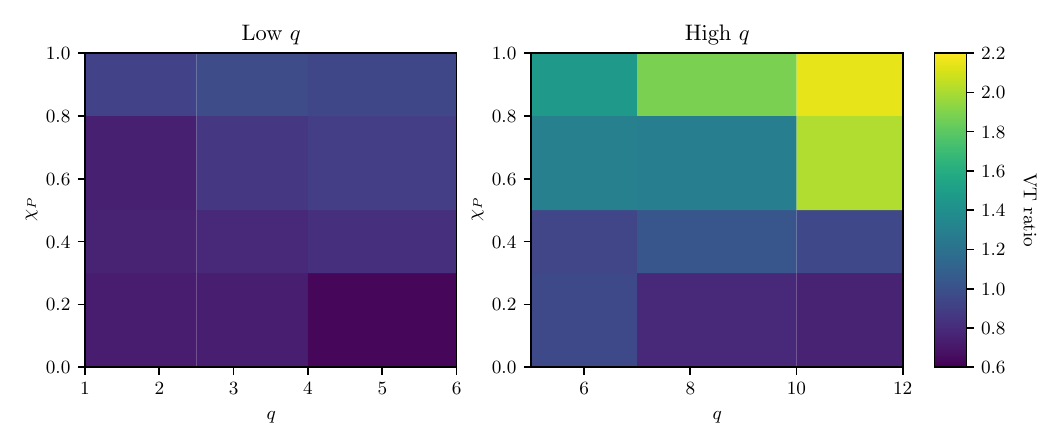}
	\caption{Sensitive volume increase for different bins in the $q-\chi_P$ space. The left panel refers to the ``Low $q$" region while the right panel to the ``High $q$" region. For each bin, we compute the VT ratio between a precessing spin search and the corresponding aligned-spin search. The VT ratio is encoded into the colorscale and is computed at a $FAR = 1/10\text{ years}$.}
	\label{fig:vt_color}
\end{figure*}

\subsection{$\xi^2$ test results} \label{sec:chisq_results}

We present in this section results from the $\xi^2$ consistency test. We will focus on the results of the ``High $q$" runs, comparing results from aligned-spin and precessing template banks.
First of all, in Fig.~\ref{fig:hist_chisq_diff} we report the difference between the $\xi^2$ values measured on the injected signals by the precessing and by the aligned-spin search.
We note that in most cases, the $\xi^2$ measured by the precessing pipeline is lower than in the aligned-spin case.
This is because the precessing templates are more similar to the injected signals, thus providing a more accurate prediction of the SNR timeseries and consequently yielding lower $\xi^2$ values. Moreover, this demonstrates the benefits of the updated $\xi^2$ expression in Eq.~\ref{eq:R_prec_HOM} over the standard expression Eq.~\ref{eq:R_std}.

In Fig.~\ref{fig:snr_chisq_plot_NP} and Fig.~\ref{fig:snr_chisq_plot_P}, we report the values of SNR and $\xi^2$ for the aligned-spin and precessing search respectively. The values refer to the LIGO-Livingston interferometer (L1).
As clear from the left panel of Fig.~\ref{fig:snr_chisq_plot_P}, for the precessing run, the $\xi^2$ is able to separate real signals from the background for SNR higher than $\simeq 9$. The result is consistent with the standard behaviour on a set of aligned-spin injections \cite{Ewing:2023qqe}.

In the Figs.~\ref{fig:snr_chisq_plot_NP}-\ref{fig:snr_chisq_plot_P} we see that both runs, aligned-spin and precessing, were able to detect the GW event GW191204\_171526 \cite{KAGRA:2021vkt}: the associated trigger is well separated from the background and hence was detected in both cases with a $\text{FAR} < \frac{1}{1000} \SI{}{yr^{-1}}$.
In the figures, one can also see that the two searches record a handful of background triggers with high SNR and low $\xi^2$, with the potential of contaminating the background of the searches. A close followup of these points revealed that such triggers were mostly caused by pairs of glitches happening very close in time with each other. This behaviour could possibly emulate the amplitude modulation typical of a precessing signal. It is worth noting that all of those potentially spurious candidates had a very high FAR, as a result of other terms of the ranking Eq.~\eqref{eq:LR} downweighting their significance.

By looking at the right panel of Fig.~\ref{fig:snr_chisq_plot_P}, we note that the injection coincidences with $\text{SNR}\gtrsim 8$ are mostly ranked with a low FAR, meaning that they are mostly recovered.
In the region characterized by $\text{SNR} \simeq 10$ and $\xi^2 \simeq 2$, the FAR slightly drops, suggesting that the $\xi^2$ test loses its efficacy. This can be probably improved by using a denser template bank (at a much larger computational cost), by implementing the un-approximated $\xi^2$ test \cite{Schmidt:2024kxy} or by improving the ranking statistics model.

In Fig.~\ref{fig:snr_chisq_plot_NP}, it is evident that, in the case of the aligned-spin bank, the $\xi^2$ values for injection signals are considerably larger than in the precessing case. Clearly, an aligned-spin template is not able to model accurately the response of a heavily precessing signal, thu measuring a high $\xi^2$ value. Consequently, injection and background triggers are not well-separated in the $\text{SNR} - \xi^2$ plane and the $\xi^2$ has dramatically downgraded performance.

It is perhaps surprising that the reduced sensitivity of the aligned-spin search is not caused by a poor SNR recovery but rather by the failure of $\xi^2$ signal consistency test. In other words, the sensitivity loss does not arise from using the ``wrong'' aligned-spin templates to filter precessing signals but rather from incorrectly labeling astrophysical triggers as non-Gaussian noise artifacts. 
This is manifest in the right panel of Fig.~\ref{fig:snr_chisq_plot_NP}, where many injections recovered with a considerably large SNR present very large $\xi^2$ values, and are consequently assigned a high FAR. This observation can have a deep impact on the planning of future large scale searches for precessing signals, as we will discuss in Sec.~\ref{sec:improvements}.

An analysis of the ``Low $q$" runs (not reported here) shows similar conclusions, even though the downranking of injection triggers due to large $\xi^2$ has less impact on the search's sensitivity, as will be discussed in the next section.


\subsection{Sensitivity improvement} \label{sec:sensitivity_results}

We finally measure the improvement in the search sensitivity brought by the use of a precessing bank both in the ``Low $q$" and in the ``High $q$" case.
As standard \cite{Tiwari:2017ndi, Tsukada:2023edh}, the sensitivity of a search is measured by space-time volume (VT), corresponding to the volume reached by a search multiplied by the observing time $T$:
\begin{equation}
	VT = T \int_0^\infty \text{d}z \, \epsilon(z, FAR) \frac{\text{d}V}{\text{d}z} \frac{1}{1+z}
\end{equation}
where $V(z)$ is the volume of the observable universe up to redshift $z$ and $\epsilon(z, FAR)$ is the efficiency of a search, defined as the fraction of injections recovered in an infinitesimal shell between $z$ and $z + \text{d}z$ with at least a given FAR.
Clearly, every estimation for $VT$ heavily depends on the distribution of injection: this motivated the use of the same injection set for the pairs of runs.

In Fig.~\ref{fig:efficiency_vs_SNR} we report the efficiency (evaluated at a $FAR = 1/10\text{ years}$) of each search as a function ``Decisive SNR", defined as the maximum between the three SNRs measured by each detector.
In the ``Low $q$" search, we note that there is not a significant difference between the aligned-spin and the precessing case. Moreover, the efficiency does not depend on the value of $\chi_P$. In this case, the improvement in the SNR recovery and in the $\xi^2$ values brought by a precessing search does not translate into an increased efficiency.
In the ``High $q$" case, the precessing search has comparable efficacy with the ``Low $q$" . On the other hand, the aligned-spin search has poorer performance, especially in recovering signals with $\chi_P>0.5$.
Thus, in the regioned covered by the ``High $q$" search, the use of precessing templates brought a substantial improvement in the efficiency of the pipeline.

In Fig.~\ref{fig:vt_vs_far}, we report the ratio $\frac{VT_\text{P}}{VT_\text{AS}}$ between the VT of by a precessing search and the VT of the corresponding aligned-spin search. The ratio is plotted as a function of FAR and we report different measurement for the low and high $\chi_P$ regions.
We note that for very high values of FAR, the precessing searches always perform better than the aligned-spin ones. This is a direct consequence of the improved SNR recovery offered by the precessing search (see also Sec.~\ref{sec:snr_results} and Fig.~\ref{fig:hist_snr_diff}).
The picture changes as soon as a lower FAR threshold is considered. Due to the increased background caused by a large number of precessing templates, the ``Low $q$" precessing search suffers from a sensitivity loss of $\sim 20\%$ as compared to its aligned-spin counterpart.
Indeed, even for highly precessing signals, with $\chi_P>0.5$, the significance of many triggers is downranked by a large amount of false positives, due to a very large number of templates which can potentially lead to high SNR triggers.
A more effective ranking statistics Eq.~\eqref{eq:LR} can mitigate this issue by providing a more robust distinction between signal and noise. This approach was successfully applied in a search targeting precessing NSBH systems~\cite{McIsaac:2023ijd}, where an updated ranking statistic was developed, improving upon its aligned-spin counterpart. Thanks to this enhancement, the authors of \cite{McIsaac:2023ijd} achieved up to a $\sim 50\%$ sensitivity improvement compared to the aligned-spin ranking statistics. Notably, their search did not reveal a decrease in performance for low values of $\chi_P$, unlike in our case.
Future studies will be needed to understand the limitations of the current ranking statistics and to develop more powerful alternatives that are better suited for detecting precessing signals.

In the ``High $q$" scenario, we still observe a decrease in sensitivity for mildly precessing signals $\chi_P<0.5$, even at low FAR. The observation above is still valid: the increased background downranks many potential candidates found by the precessing search.
On the other hand, for $\chi_P>0.5$ we observe a $\sim 50\%$ increase of the volume reached by the precessing search. The increase remains stable with the FAR.
These considerations are consistent with our observations regarding the measured search efficiency Fig.~\ref{fig:efficiency_vs_SNR}.

It is remarkable, if not surprising, that in the ``High $q$" case an improvement in sensitivity was obtained {\it despite} the fact that the precessing bank was generated with $0.9$ target minimal match. Moreover, the ``Low $q$" aligned-spin bank, provides an excellent recovery of precessing injections, despite the fact that more than $10\%$ of the injections considered have a match of $0.9$ and lower.
The two coupled observations seems to indicate that also future searches for precessing signals can employ $0.9$ target minimal match template banks. This has the potential of reducing by orders of magnitude the computational cost as well as the background of the search.

In Fig.~\ref{fig:vt_color}, the VT ratio is computed for different bins in the $q - \chi_P$ plane for a fixed FAR of $1/10\text{ years}$. Results in in the left and right panels refer to the ``Low $q$" and ``High $q$" respectively and they corroborate what observed above.
In the ``Low $q$" case, we see little variation of the sensitive volume across the space, and for low values of $q$ and $\chi_P$, we observe up to a $40\%$ decrease in sensitivity due to the largely increased background. On the other hand, in the ``High $q$" scenario we note that VT ratio can be as high as $100\%$, depending on the values of $q$ and $\chi_P$.
As already noted before, for low $\chi_P$ the precessing search does not bring any improvement. On the other, for $\chi_P>0.5$, the sensitivity rapidly grows as $\chi_P$ and $q$ grow, reaching a $120\%$ gain in the most extreme case.

Our results suggests that targeting more asymmetric systems, such as a NSBH system, could potentially lead to an even larger sensitivity improvement. However as noted in \cite{Schmidt:2023gzj}, a search in such region struggles with an increased size of the parameter space. Indeed, a template bank in the NSBH region has hundreds of millions of templates and probably a different approach to the search might be needed. In this respect, the work presented in \cite{McIsaac:2023ijd} is a promising avenue to target the precessing NSBH parameter space, due its lower computational cost.



\section{Outline of future improvements} \label{sec:improvements}

Up to this point, we have adapted the GstLAL pipeline to handle precessing templates by modifying the SNR computation routine and making minimal edits to the $\xi^2$ test, as outlined in Sec.~\ref{sec:methods}.
However, other components of the pipeline have remained unchanged. Although our modifications have successfully enhanced the pipeline sensitivity for precessing signals, there might still be room for improvement in the pipeline's overall performance. Below, we summarize several potential directions for future developments.

Firstly, it's important to acknowledge that the $\xi^2$ test implemented in this work represents an approximate version of the test proposed in \cite{Schmidt:2024kxy}. Implementing the full $\xi^2$ test would necessitate extensive changes to the pipeline, which are beyond the scope of the present paper.
In \cite{Schmidt:2024kxy}, an injection campaign in both Gaussian and real noise was conducted, suggesting that for signals with $\text{SNR}\lesssim 20$, the approximation employed in our implementation of $\xi^2$ does not introduce significant bias. However, for high SNR signals, the bias introduced by our approximation becomes more significant.
We note that the study in \cite{Schmidt:2024kxy} did not account for the discreteness of the template bank, which can potentially degrade the performance of the currently implemented $\xi^2$ test Eq.~\eqref{eq:R_prec_HOM} in a real search scenario. Investigating this aspect is deferred to future work and is essential for determining whether modifying the $\xi^2$ test can lead to a substantial increase in the performance of a precessing search.

In our searches, we have not employed the ``bank $\xi^2$", introduced in \cite{Tsukada:2023edh}.
The ``bank $\xi^2$" is an alternative signal consistency test, which checks for a consistent response among several templates within the same SVD bin. It is used as an additional test to the ``original" $\xi^2$ and it has been shown to provide an improvement in the high mass region.
Future work should assess the efficacy of the ``bank $\xi^2$'' applied to precessing signals and possibly deploy it in a real search.

As previously discussed, the ranking statistics given by Eq.~\eqref{eq:LR} may also need adaptation for precessing templates.
The key LR term requiring modification is the ``coherence" term, as highlighted in Sec.~\ref{sec:LR}. In the precessing case, the measured phase of a signal $\phi$ loses the straightforward interpretation it has for an aligned-spin signal. Since this information has been crucial in producing the ``coherence" term of the ranking statistics, it might not be directly suitable for a precessing template.
Deriving an analytical expression that generalizes the ``coherence" term to precession is likely to be challenging. In this context, utilizing an unsupervised Machine Learning model could prove to be an effective solution, allowing for a more flexible and data-driven adaptation of the ranking statistics to precessing templates.

Besides the ``coherence" term, other factors of the ranking statistics retain the same physical meaning they have for an aligned-spin search. For this reason, in principle, they do not require modification. However, their predictive power depend on some hyperparameters, which were tuned by only considering aligned-spin searches. Tuning such hyperparameters on a set of precessing injections can offer an increased performance for the search.

We observed in Sec.~\ref{sec:chisq_results} that the primary loss in sensitivity by the aligned-spin ``High $q$" precessing search arises from the degraded performance of the $\xi^2$ test. This fact can have a profound impact on the search for precessing signals. Indeed, as the SNR recovery is not crucial for performance, it should be possible to deploy a template bank with a very low target minimal match ($0.8$, or even smaller). The reduced size of the resulting bank will have two favourable implications: (i) it reduces the computational cost and (ii) it reduces the amount of false positives, possibly improving the sensitivity. Of course, a more sparse template bank will also downgrade the performance of the $\xi^2$ test and future work should explore the trade off between the benefits of a smaller template bank and the downsides of a less effective signal consistency test.
With this strategy, it should become feasible to carry on a precessing search in the NSBH region of the parameter space, where the precession content of a signal is expected to be very strong.

Finally, it is important to stress that the changes we made to the pipeline are fully applicable to systems with HOM content.
Similar to the approach taken for precession in this work, future efforts will evaluate the pipeline's performance in the presence of HOM signals and assess potential sensitivity improvements. Once results from an HOM search become available, the considerations outlined above regarding possible further pipeline upgrades remain relevant.

\section{Final remarks} \label{sec:conclusion}

This paper details the modifications made to the GstLAL pipeline to enable a systematic search for GW signals originating from precessing BBH systems.
To filter the data, we incorporated the search statistics outlined in \cite{Capano:2013raa, Schmidt:2014iyl, Harry:2017weg} by employing the ``orthogonalized" template Eq.~\eqref{eq:h_orth}. This is also discussed in App.~\ref{app:orthogonalization}. To account for the effects of this modified template response, we made updates to the $\xi^2$ consistency test. The test now implemented is an approximation to the one proposed in \cite{Schmidt:2024kxy}.

To assess the effectiveness of our enhancements, we generated two large precessing template banks to search for precessing signals in two distinct regions of the BBH parameter space.
These template banks were employed to search one week of publicly available data from the third observing run, spanning between GPS times $1259423400$ and $1260081799$. The results obtained from these searches were then compared with two identical searches conducted using an aligned-spin template bank, which covered the same mass range.

As discussed in Sec.~\ref{sec:search_results}, searching the data with precessing templates results in a slight increase of a signal recovered SNR and to a substantial improvement in the $\xi^2$ test performance.
However, this is achieved at the price of a large increase in the background triggers, due to a template bank being up to two orders of magnitude larger than in the corresponding aligned-spin case.
We obtain a substantial improvement of the sensitivity of the pipeline only when searching strongly precessing systems, characterized by a high mass ratio $q \gtrsim 5$ and highly precessing spin $\chi_P \geq 0.5$. For less strongly precessing systems, we observe a slight decrease of the pipeline sensitivity, as the increased background removes any positive effect of the improved SNR recovery and of the more accurate signal consistency test.

In Sec.~\ref{sec:improvements}, we outlined a few possible future upgrades of the pipeline, including an updated ranking statistics Eq.~\eqref{eq:LR} and the implementation of the ``exact" signal consistency test \cite{Schmidt:2024kxy}.
Moreover, the pipeline developed for precessing signals can be straightforwardly applied to the search of signals with HOM content. Future work should aim at assessing the sensitivity brought by a similar search.

Our results lead to two significant conclusions that can inform the planning of future precessing searches.
The first observation is technical: we find that employing a template bank with a target minimal match of $0.9$ is sufficient to achieve a substantial sensitivity improvement for highly precessing signals. As discussed in Sec.~\ref{sec:chisq_results}, this is due to the fact that the primary sensitivity loss of an aligned-spin search targeting the precessing region is not due to a lack of recovered SNR but rather by a worsening of the $\xi^2$ value. The observation is crucial for reducing the size of the template bank, consequently lowering computational costs and the associated background in a search.
The second observation pertains to the targeted parameter space. Our results suggest that, to attain any sensitivity improvement, a precessing search should focus solely on asymmetric and heavily precessing systems, characterized by $q \gtrsim 5$ and $\chi_P\gtrsim 0.5$. Current aligned-spin searches have limited efficacy in such an extreme region, and indeed, signals in this range have not been detected by past aligned-spin searches.

Our work provides the community with the tools to search for precessing signals. Moreover, our investigations enable the community to select an appropriate parameter space to target with a precessing search, where a large sensitivity improvement, and possibly new detections, are within reach.

        \begin{acknowledgments}
        We thank Melissa Lopez Portilla for sharing her knowledge on glitches and detector characterization.
		S.S. is supported by the research program of the Netherlands Organisation for Scientific Research (NWO). S.C. is supported by the National Science Foundation under Grant No. PHY-2309332. The authors are grateful for computational resources provided by the LIGO Laboratory and supported by National Science Foundation Grants No. PHY-0757058 and No. PHY-0823459. This material is based upon work supported by NSF’s LIGO Laboratory which is a major facility fully funded by the National Science Foundation.
		LIGO was constructed by the California Institute of Technology and Massachusetts Institute of Technology with funding from the National Science Foundation (NSF) and operates under cooperative Agreement No. PHY-1764464. The authors are grateful for computational resources provided by the Pennsylvania State University’s Institute for Computational and Data Sciences (ICDS) and the University of Wisconsin Milwaukee Nemo and support by NSF Grant No. PHY-2011865, No. NSF OAC-2103662, No. NSF PHY-1626190, No. NSF PHY-1700765, No. NSF PHY-2207728, No. NSF PHY-2207594 and No. PHY-2309332 as well as by the research program of the Netherlands Organization for Scientific Research (NWO).
		This paper carries LIGO Document No. LIGO-P2400044. This research has made use of data or software obtained from the Gravitational Wave Open Science Center (gwosc.org), a service of LIGO Laboratory, the LIGO Scientific Collaboration, the Virgo Collaboration, and KAGRA. LIGO Laboratory and Advanced LIGO are funded by the United States National Science Foundation (NSF) as well as the Science and Technology Facilities Council (STFC) of the United Kingdom, the Max-Planck-Society (MPS), and the State of Niedersachsen/Germany for support of the construction of Advanced LIGO and construction and operation of the GEO600 detector. Additional support for Advanced LIGO was provided by the Australian Research Council. Virgo is funded, through the European Gravitational Observatory (EGO), by the French Centre National de Recherche Scientifique (CNRS), the Italian Istituto Nazionale di Fisica Nucleare (INFN) and the Dutch Nikhef, with contributions by institutions from Belgium, Germany, Greece, Hungary, Ireland, Japan, Monaco, Poland, Portugal, Spain. KAGRA is supported by Ministry of Education, Culture, Sports, Science and Technology (MEXT), Japan Society for the Promotion of Science (JSPS) in Japan; National Research Foundation (NRF) and Ministry of Science and ICT (MSIT) in Korea; Academia Sinica (AS) and National Science and Technology Council (NSTC) in Taiwan.
        \end{acknowledgments}

\appendix

\section{Template orthogonalization}\label{app:orthogonalization}

In this appendix we show that the search statistics Eq.~\eqref{eq:symphony_snr}
is equivalent to:
\begin{equation}\label{eq:symphony_snr_orth}
	\max \Lambda = \rescalar{s}{\hat{h}_+}^2 + \rescalar{s}{\hat{h}_\perp}^2
\end{equation}
where the ``orthogonalized" template $\hat{h}_\perp$ was introduced in Eq.~\eqref{eq:h_orth}:
\begin{equation}
	\hat{h}_\perp = \frac{\hat{h}_\times - \hat{h}_{+\times} \hat{h}_+}{\sqrt{1- \hat{h}^2_{+\times}}}
\end{equation}
Eq.~\eqref{eq:symphony_snr_orth} is more computationally convenient than Eq.~\eqref{eq:symphony_snr}. Moreover it is well suited to be incorporated into the GstLAL pipeline, since it only involves a minimal change in the template generation routine and no changes in the filtering routine.

By defining $s_{+/\times} =  \rescalar{s}{\hat{h}_{+/\times}}$ and $\hat{h}_{+\times} = \rescalar{\hat{h}_+}{\hat{h}_\times}$, we trivially obtain:
\begin{align*}\label{eq:orthogonalization_computation}
	\max \Lambda &= \rescalar{s}{\hat{h}_+}^2 + \frac{1}{1- \hat{h}^2_{+\times}} \rescalar{s}{\hat{h}_\times - \hat{h}_{+\times} \hat{h}_+}^2 \\
				 &= s^2_+ + \frac{\hat{h}_{+\times}}{1- \hat{h}^2_{+\times}} s^2_+ + \frac{1}{1- \hat{h}^2_{+\times}} s^2_\times
				 	- 2 \frac{\hat{h}^2_{+\times}}{1- \hat{h}^2_{+\times}} s_+ s_\times \\
				 &= \frac{1}{1- \hat{h}^2_{+\times}} \left\{ s^2_+ + s^2_\times -2 \hat{h}_{+\times} s_+ s_\times \right\}	
\end{align*}
In the last line, we can easily recognize Eq.~\eqref{eq:symphony_snr}.
This proves the validity of Eq.~\eqref{eq:complex_snr_symphony} for filtering the data with a precessing and/or HOM template.

	\bibliography{biblio.bib}

\begin{thebibliography}{100}

\bibitem{Apostolatos:1994mx}
T.~A. Apostolatos, C.~Cutler, G.~J. Sussman, and K.~S. Thorne, ``{Spin induced
  orbital precession and its modulation of the gravitational wave forms from
  merging binaries},'' {\em Phys. Rev. D}, vol.~49, pp.~6274--6297, 1994.

\bibitem{Kidder:1992fr}
L.~E. Kidder, C.~M. Will, and A.~G. Wiseman, ``{Spin effects in the inspiral of
  coalescing compact binaries},'' {\em Phys. Rev. D}, vol.~47, no.~10,
  pp.~R4183--R4187, 1993.

\bibitem{Kidder:1995zr}
L.~E. Kidder, ``{Coalescing binary systems of compact objects to postNewtonian
  5/2 order. 5. Spin effects},'' {\em Phys. Rev. D}, vol.~52, pp.~821--847,
  1995.

\bibitem{Buonanno:2002fy}
A.~Buonanno, Y.-b. Chen, and M.~Vallisneri, ``{Detecting gravitational waves
  from precessing binaries of spinning compact objects: Adiabatic limit},''
  {\em Phys. Rev. D}, vol.~67, p.~104025, 2003.
\newblock [Erratum: Phys.Rev.D 74, 029904 (2006)].

\bibitem{Campanelli:2006fy}
M.~Campanelli, C.~O. Lousto, Y.~Zlochower, B.~Krishnan, and D.~Merritt, ``{Spin
  Flips and Precession in Black-Hole-Binary Mergers},'' {\em Phys. Rev. D},
  vol.~75, p.~064030, 2007.

\bibitem{Racine:2008kj}
E.~Racine, A.~Buonanno, and L.~E. Kidder, ``{Recoil velocity at 2PN order for
  spinning black hole binaries},'' {\em Phys. Rev. D}, vol.~80, p.~044010,
  2009.

\bibitem{Bohe:2012mr}
A.~Bohe, S.~Marsat, G.~Faye, and L.~Blanchet, ``{Next-to-next-to-leading order
  spin-orbit effects in the near-zone metric and precession equations of
  compact binaries},'' {\em Class. Quant. Grav.}, vol.~30, p.~075017, 2013.

\bibitem{Bohe:2015ana}
A.~Boh\'e, G.~Faye, S.~Marsat, and E.~K. Porter, ``{Quadratic-in-spin effects
  in the orbital dynamics and gravitational-wave energy flux of compact
  binaries at the 3PN order},'' {\em Class. Quant. Grav.}, vol.~32, no.~19,
  p.~195010, 2015.

\bibitem{Pratten:2020ceb}
G.~Pratten {\em et~al.}, ``{Computationally efficient models for the dominant
  and subdominant harmonic modes of precessing binary black holes},'' {\em
  Phys. Rev. D}, vol.~103, no.~10, p.~104056, 2021.

\bibitem{Estelles:2020osj}
H.~Estell\'es, A.~Ramos-Buades, S.~Husa, C.~Garc\'\i{}a-Quir\'os, M.~Colleoni,
  L.~Haegel, and R.~Jaume, ``{Phenomenological time domain model for dominant
  quadrupole gravitational wave signal of coalescing binary black holes},''
  {\em Phys. Rev. D}, vol.~103, no.~12, p.~124060, 2021.

\bibitem{Akcay:2020qrj}
S.~Akcay, R.~Gamba, and S.~Bernuzzi, ``{Hybrid post-Newtonian
  effective-one-body scheme for spin-precessing compact-binary waveforms up to
  merger},'' {\em Phys. Rev. D}, vol.~103, no.~2, p.~024014, 2021.

\bibitem{Hamilton:2021pkf}
E.~Hamilton, L.~London, J.~E. Thompson, E.~Fauchon-Jones, M.~Hannam,
  C.~Kalaghatgi, S.~Khan, F.~Pannarale, and A.~Vano-Vinuales, ``{Model of
  gravitational waves from precessing black-hole binaries through merger and
  ringdown},'' {\em Phys. Rev. D}, vol.~104, no.~12, p.~124027, 2021.

\bibitem{Gamba:2021ydi}
R.~Gamba, S.~Ak\c{c}ay, S.~Bernuzzi, and J.~Williams, ``{Effective-one-body
  waveforms for precessing coalescing compact binaries with post-Newtonian
  twist},'' {\em Phys. Rev. D}, vol.~106, no.~2, p.~024020, 2022.

\bibitem{Ramos-Buades:2023ehm}
A.~Ramos-Buades, A.~Buonanno, H.~Estell\'es, M.~Khalil, D.~P. Mihaylov,
  S.~Ossokine, L.~Pompili, and M.~Shiferaw, ``{SEOBNRv5PHM: Next generation of
  accurate and efficient multipolar precessing-spin effective-one-body
  waveforms for binary black holes},'' 3 2023.

\bibitem{LIGOScientific:2016vbw}
B.~P. Abbott {\em et~al.}, ``{GW150914: First results from the search for
  binary black hole coalescence with Advanced LIGO},'' {\em Phys. Rev. D},
  vol.~93, no.~12, p.~122003, 2016.

\bibitem{LIGOScientific:2016vlm}
B.~P. Abbott {\em et~al.}, ``{Properties of the Binary Black Hole Merger
  GW150914},'' {\em Phys. Rev. Lett.}, vol.~116, no.~24, p.~241102, 2016.

\bibitem{LIGOScientific:2016gtq}
B.~P. Abbott {\em et~al.}, ``{Characterization of transient noise in Advanced
  LIGO relevant to gravitational wave signal GW150914},'' {\em Class. Quant.
  Grav.}, vol.~33, no.~13, p.~134001, 2016.

\bibitem{LIGOScientific:2014pky}
J.~Aasi {\em et~al.}, ``{Advanced LIGO},'' {\em Class. Quant. Grav.}, vol.~32,
  p.~074001, 2015.

\bibitem{VIRGO:2014yos}
F.~Acernese {\em et~al.}, ``{Advanced Virgo: a second-generation
  interferometric gravitational wave detector},'' {\em Class. Quant. Grav.},
  vol.~32, no.~2, p.~024001, 2015.

\bibitem{KAGRA:2020tym}
T.~Akutsu {\em et~al.}, ``{Overview of KAGRA: Detector design and construction
  history},'' {\em PTEP}, vol.~2021, no.~5, p.~05A101, 2021.

\bibitem{LIGOScientific:2018mvr}
B.~P. Abbott {\em et~al.}, ``{GWTC-1: A Gravitational-Wave Transient Catalog of
  Compact Binary Mergers Observed by LIGO and Virgo during the First and Second
  Observing Runs},'' {\em Phys. Rev. X}, vol.~9, no.~3, p.~031040, 2019.

\bibitem{LIGOScientific:2020ibl}
R.~Abbott {\em et~al.}, ``{GWTC-2: Compact Binary Coalescences Observed by LIGO
  and Virgo During the First Half of the Third Observing Run},'' {\em Phys.
  Rev. X}, vol.~11, p.~021053, 2021.

\bibitem{LIGOScientific:2021usb}
R.~Abbott {\em et~al.}, ``{GWTC-2.1: Deep extended catalog of compact binary
  coalescences observed by LIGO and Virgo during the first half of the third
  observing run},'' {\em Phys. Rev. D}, vol.~109, no.~2, p.~022001, 2024.

\bibitem{KAGRA:2021vkt}
R.~Abbott {\em et~al.}, ``{GWTC-3: Compact Binary Coalescences Observed by LIGO
  and Virgo during the Second Part of the Third Observing Run},'' {\em Phys.
  Rev. X}, vol.~13, no.~4, p.~041039, 2023.

\bibitem{LIGOScientific:2020kqk}
R.~Abbott {\em et~al.}, ``{Population Properties of Compact Objects from the
  Second LIGO-Virgo Gravitational-Wave Transient Catalog},'' {\em Astrophys. J.
  Lett.}, vol.~913, no.~1, p.~L7, 2021.

\bibitem{KAGRA:2021duu}
R.~Abbott {\em et~al.}, ``{Population of Merging Compact Binaries Inferred
  Using Gravitational Waves through GWTC-3},'' {\em Phys. Rev. X}, vol.~13,
  no.~1, p.~011048, 2023.

\bibitem{Schmidt:2014iyl}
P.~Schmidt, F.~Ohme, and M.~Hannam, ``{Towards models of gravitational
  waveforms from generic binaries II: Modelling precession effects with a
  single effective precession parameter},'' {\em Phys. Rev. D}, vol.~91, no.~2,
  p.~024043, 2015.

\bibitem{LIGOScientific:2020iuh}
R.~Abbott {\em et~al.}, ``{GW190521: A Binary Black Hole Merger with a Total
  Mass of $150 M_{\odot}$},'' {\em Phys. Rev. Lett.}, vol.~125, no.~10,
  p.~101102, 2020.

\bibitem{Zhang:2023fpp}
R.~C. Zhang, G.~Fragione, C.~Kimball, and V.~Kalogera, ``{On the Likely
  Dynamical Origin of GW191109 and Binary Black Hole Mergers with Negative
  Effective Spin},'' {\em Astrophys. J.}, vol.~954, no.~1, p.~23, 2023.

\bibitem{Hannam:2021pit}
M.~Hannam {\em et~al.}, ``{General-relativistic precession in a black-hole
  binary},'' {\em Nature}, vol.~610, no.~7933, pp.~652--655, 2022.

\bibitem{Payne:2022spz}
E.~Payne, S.~Hourihane, J.~Golomb, R.~Udall, R.~Udall, D.~Davis, and
  K.~Chatziioannou, ``{Curious case of GW200129: Interplay between
  spin-precession inference and data-quality issues},'' {\em Phys. Rev. D},
  vol.~106, no.~10, p.~104017, 2022.

\bibitem{Macas:2023wiw}
R.~Macas, A.~Lundgren, and G.~Ashton, ``{Revisiting GW200129 with machine
  learning noise mitigation: it is (still) precessing},'' 11 2023.

\bibitem{Vitale:2018wlg}
S.~Vitale and H.-Y. Chen, ``{Measuring the Hubble constant with neutron star
  black hole mergers},'' {\em Phys. Rev. Lett.}, vol.~121, no.~2, p.~021303,
  2018.

\bibitem{Yun:2023ygz}
Q.~Yun, W.-B. Han, Q.~Hu, and H.~Xu, ``{Precessing binary black holes as better
  dark sirens},'' {\em Mon. Not. Roy. Astron. Soc.}, vol.~527, no.~1,
  pp.~L60--L65, 2023.

\bibitem{Farr:2017gtv}
B.~Farr, D.~E. Holz, and W.~M. Farr, ``{Using Spin to Understand the Formation
  of LIGO and Virgo\textquoteright{}s Black Holes},'' {\em Astrophys. J.
  Lett.}, vol.~854, no.~1, p.~L9, 2018.

\bibitem{Farr:2017uvj}
W.~M. Farr, S.~Stevenson, M.~Coleman~Miller, I.~Mandel, B.~Farr, and
  A.~Vecchio, ``{Distinguishing Spin-Aligned and Isotropic Black Hole
  Populations With Gravitational Waves},'' {\em Nature}, vol.~548, p.~426,
  2017.

\bibitem{Johnson-McDaniel:2021rvv}
N.~K. Johnson-McDaniel, S.~Kulkarni, and A.~Gupta, ``{Inferring spin tilts at
  formation from gravitational wave observations of binary black holes:
  Interfacing precession-averaged and orbit-averaged spin evolution},'' {\em
  Phys. Rev. D}, vol.~106, no.~2, p.~023001, 2022.

\bibitem{Vitale:2022dpa}
S.~Vitale, S.~Biscoveanu, and C.~Talbot, ``{Spin it as you like: The (lack of
  a) measurement of the spin tilt distribution with LIGO-Virgo-KAGRA binary
  black holes},'' {\em Astron. Astrophys.}, vol.~668, p.~L2, 2022.

\bibitem{Mapelli:2021taw}
M.~Mapelli, {\em {Formation Channels of Single and Binary Stellar-Mass Black
  Holes}}.
\newblock 2021.

\bibitem{Klimenko:2008fu}
S.~Klimenko, I.~Yakushin, A.~Mercer, and G.~Mitselmakher, ``{Coherent method
  for detection of gravitational wave bursts},'' {\em Class. Quant. Grav.},
  vol.~25, p.~114029, 2008.

\bibitem{Klimenko:2011hz}
S.~Klimenko, G.~Vedovato, M.~Drago, G.~Mazzolo, G.~Mitselmakher, C.~Pankow,
  G.~Prodi, V.~Re, F.~Salemi, and I.~Yakushin, ``{Localization of gravitational
  wave sources with networks of advanced detectors},'' {\em Phys. Rev. D},
  vol.~83, p.~102001, 2011.

\bibitem{Klimenko:2015ypf}
S.~Klimenko {\em et~al.}, ``{Method for detection and reconstruction of
  gravitational wave transients with networks of advanced detectors},'' {\em
  Phys. Rev. D}, vol.~93, no.~4, p.~042004, 2016.

\bibitem{Drago:2020kic}
M.~Drago {\em et~al.}, ``{Coherent WaveBurst, a pipeline for unmodeled
  gravitational-wave data analysis},'' 6 2020.

\bibitem{Allen:2005fk}
B.~Allen, W.~G. Anderson, P.~R. Brady, D.~A. Brown, and J.~D.~E. Creighton,
  ``{FINDCHIRP: An Algorithm for detection of gravitational waves from
  inspiraling compact binaries},'' {\em Phys. Rev. D}, vol.~85, p.~122006,
  2012.

\bibitem{Nitz:2021zwj}
A.~H. Nitz, S.~Kumar, Y.-F. Wang, S.~Kastha, S.~Wu, M.~Sch\"afer, R.~Dhurkunde,
  and C.~D. Capano, ``{4-OGC: Catalog of Gravitational Waves from Compact
  Binary Mergers},'' {\em Astrophys. J.}, vol.~946, no.~2, p.~59, 2023.

\bibitem{Nitz:2021uxj}
A.~H. Nitz, C.~D. Capano, S.~Kumar, Y.-F. Wang, S.~Kastha, M.~Sch\"afer,
  R.~Dhurkunde, and M.~Cabero, ``{3-OGC: Catalog of Gravitational Waves from
  Compact-binary Mergers},'' {\em Astrophys. J.}, vol.~922, no.~1, p.~76, 2021.

\bibitem{Olsen:2022pin}
S.~Olsen, T.~Venumadhav, J.~Mushkin, J.~Roulet, B.~Zackay, and M.~Zaldarriaga,
  ``{New binary black hole mergers in the LIGO-Virgo O3a data},'' {\em Phys.
  Rev. D}, vol.~106, no.~4, p.~043009, 2022.

\bibitem{Mehta:2023zlk}
A.~K. Mehta, S.~Olsen, D.~Wadekar, J.~Roulet, T.~Venumadhav, J.~Mushkin,
  B.~Zackay, and M.~Zaldarriaga, ``{New binary black hole mergers in the
  LIGO-Virgo O3b data},'' 11 2023.

\bibitem{Harry:2013tca}
I.~W. Harry, A.~H. Nitz, D.~A. Brown, A.~P. Lundgren, E.~Ochsner, and
  D.~Keppel, ``{Investigating the effect of precession on searches for
  neutron-star-black-hole binaries with Advanced LIGO},'' {\em Phys. Rev. D},
  vol.~89, no.~2, p.~024010, 2014.

\bibitem{CalderonBustillo:2016rlt}
J.~Calder\'on~Bustillo, P.~Laguna, and D.~Shoemaker, ``{Detectability of
  gravitational waves from binary black holes: Impact of precession and higher
  modes},'' {\em Phys. Rev. D}, vol.~95, no.~10, p.~104038, 2017.

\bibitem{Dhurkunde:2022abc}
R.~Dhurkunde and A.~H. Nitz, ``Sensitivity of spin-aligned searches for neutron
  star-black hole systems using future detectors,'' {\em Phys. Rev. D},
  vol.~106, p.~103035, Nov 2022.

\bibitem{Messick:2016aqy}
C.~Messick {\em et~al.}, ``{Analysis Framework for the Prompt Discovery of
  Compact Binary Mergers in Gravitational-wave Data},'' {\em Phys. Rev. D},
  vol.~95, no.~4, p.~042001, 2017.

\bibitem{Sachdev:2019vvd}
S.~Sachdev {\em et~al.}, ``{The GstLAL Search Analysis Methods for Compact
  Binary Mergers in Advanced LIGO's Second and Advanced Virgo's First Observing
  Runs},'' 1 2019.

\bibitem{Hanna:2019ezx}
C.~Hanna {\em et~al.}, ``{Fast evaluation of multidetector consistency for
  real-time gravitational wave searches},'' {\em Phys. Rev. D}, vol.~101,
  no.~2, p.~022003, 2020.

\bibitem{cannon2020gstlal}
K.~Cannon, S.~Caudill, C.~Chan, B.~Cousins, J.~D.~E. Creighton, B.~Ewing,
  H.~Fong, P.~Godwin, C.~Hanna, S.~Hooper, R.~Huxford, R.~Magee, D.~Meacher,
  C.~Messick, S.~Morisaki, D.~Mukherjee, H.~Ohta, A.~Pace, S.~Privitera,
  I.~de~Ruiter, S.~Sachdev, L.~Singer, D.~Singh, R.~Tapia, L.~Tsukada,
  D.~Tsuna, T.~Tsutsui, K.~Ueno, A.~Viets, L.~Wade, and M.~Wade, ``Gstlal: A
  software framework for gravitational wave discovery,'' 2020.

\bibitem{Ewing:2023qqe}
B.~Ewing {\em et~al.}, ``{Performance of the low-latency GstLAL inspiral search
  towards LIGO, Virgo, and KAGRA's fourth observing run},'' 5 2023.

\bibitem{Tsukada:2023edh}
L.~Tsukada {\em et~al.} {\em Phys. Rev. D}, vol.~108, no.~4, p.~043004, 2023.

\bibitem{DalCanton:2014hxh}
T.~Dal~Canton {\em et~al.}, ``{Implementing a search for aligned-spin neutron
  star-black hole systems with advanced ground based gravitational wave
  detectors},'' {\em Phys. Rev. D}, vol.~90, no.~8, p.~082004, 2014.

\bibitem{Usman:2015kfa}
S.~A. Usman {\em et~al.}, ``{The PyCBC search for gravitational waves from
  compact binary coalescence},'' {\em Class. Quant. Grav.}, vol.~33, no.~21,
  p.~215004, 2016.

\bibitem{Nitz:2017svb}
A.~H. Nitz, T.~Dent, T.~Dal~Canton, S.~Fairhurst, and D.~A. Brown, ``{Detecting
  binary compact-object mergers with gravitational waves: Understanding and
  Improving the sensitivity of the PyCBC search},'' {\em Astrophys. J.},
  vol.~849, no.~2, p.~118, 2017.

\bibitem{Davies:2020tsx}
G.~S. Davies, T.~Dent, M.~T\'apai, I.~Harry, C.~McIsaac, and A.~H. Nitz,
  ``{Extending the PyCBC search for gravitational waves from compact binary
  mergers to a global network},'' {\em Phys. Rev. D}, vol.~102, no.~2,
  p.~022004, 2020.

\bibitem{Adams:2015ulm}
T.~Adams, D.~Buskulic, V.~Germain, G.~M. Guidi, F.~Marion, M.~Montani,
  B.~Mours, F.~Piergiovanni, and G.~Wang, ``{Low-latency analysis pipeline for
  compact binary coalescences in the advanced gravitational wave detector
  era},'' {\em Class. Quant. Grav.}, vol.~33, no.~17, p.~175012, 2016.

\bibitem{Aubin:2020goo}
F.~Aubin {\em et~al.}, ``{The MBTA pipeline for detecting compact binary
  coalescences in the third LIGO\textendash{}Virgo observing run},'' {\em
  Class. Quant. Grav.}, vol.~38, no.~9, p.~095004, 2021.

\bibitem{Luan:2011qx}
J.~Luan, S.~Hooper, L.~Wen, and Y.~Chen, ``{Towards low-latency real-time
  detection of gravitational waves from compact binary coalescences in the era
  of advanced detectors},'' {\em Phys. Rev. D}, vol.~85, p.~102002, 2012.

\bibitem{Chu:2017ovg}
Q.~Chu, {\em {Low-latency detection and localization of gravitational waves
  from compact binary coalescences}}.
\newblock PhD thesis, Western Australia U., 2017.

\bibitem{Chu:2020pjv}
Q.~Chu {\em et~al.}, ``{SPIIR online coherent pipeline to search for
  gravitational waves from compact binary coalescences},'' {\em Phys. Rev. D},
  vol.~105, no.~2, p.~024023, 2022.

\bibitem{Privitera:2013xza}
S.~Privitera, S.~R.~P. Mohapatra, P.~Ajith, K.~Cannon, N.~Fotopoulos, M.~A.
  Frei, C.~Hanna, A.~J. Weinstein, and J.~T. Whelan, ``{Improving the
  sensitivity of a search for coalescing binary black holes with nonprecessing
  spins in gravitational wave data},'' {\em Phys. Rev. D}, vol.~89, no.~2,
  p.~024003, 2014.

\bibitem{Capano:2016dsf}
C.~Capano, I.~Harry, S.~Privitera, and A.~Buonanno, ``{Implementing a search
  for gravitational waves from binary black holes with nonprecessing spin},''
  {\em Phys. Rev. D}, vol.~93, no.~12, p.~124007, 2016.

\bibitem{Venumadhav:2019tad}
T.~Venumadhav, B.~Zackay, J.~Roulet, L.~Dai, and M.~Zaldarriaga, ``{New search
  pipeline for compact binary mergers: Results for binary black holes in the
  first observing run of Advanced LIGO},'' {\em Phys. Rev. D}, vol.~100, no.~2,
  p.~023011, 2019.

\bibitem{Sathyaprakash:1991mt}
B.~S. Sathyaprakash and S.~V. Dhurandhar, ``{Choice of filters for the
  detection of gravitational waves from coalescing binaries},'' {\em Phys. Rev.
  D}, vol.~44, pp.~3819--3834, 1991.

\bibitem{Dhurandhar:1992mw}
S.~V. Dhurandhar and B.~S. Sathyaprakash, ``{Choice of filters for the
  detection of gravitational waves from coalescing binaries. 2. Detection in
  colored noise},'' {\em Phys. Rev. D}, vol.~49, pp.~1707--1722, 1994.

\bibitem{Owen:1998dk}
B.~J. Owen and B.~S. Sathyaprakash, ``{Matched filtering of gravitational waves
  from inspiraling compact binaries: Computational cost and template
  placement},'' {\em Phys. Rev. D}, vol.~60, p.~022002, 1999.

\bibitem{Babak:2006ty}
S.~Babak, R.~Balasubramanian, D.~Churches, T.~Cokelaer, and B.~S.
  Sathyaprakash, ``{A Template bank to search for gravitational waves from
  inspiralling compact binaries. I. Physical models},'' {\em Class. Quant.
  Grav.}, vol.~23, pp.~5477--5504, 2006.

\bibitem{Cokelaer:2007mv}
T.~Cokelaer, ``{A Template bank to search for gravitational waves from
  inspiralling compact binaries. II. Phenomenological model},'' {\em Class.
  Quant. Grav.}, vol.~24, pp.~6227--6242, 2007.

\bibitem{Harry:2009ea}
I.~W. Harry, B.~Allen, and B.~S. Sathyaprakash, ``{A Stochastic template
  placement algorithm for gravitational wave data analysis},'' {\em Phys. Rev.
  D}, vol.~80, p.~104014, 2009.

\bibitem{Ajith:2012mn}
P.~Ajith, N.~Fotopoulos, S.~Privitera, A.~Neunzert, and A.~J. Weinstein,
  ``{Effectual template bank for the detection of gravitational waves from
  inspiralling compact binaries with generic spins},'' {\em Phys. Rev. D},
  vol.~89, no.~8, p.~084041, 2014.

\bibitem{Roy:2017oul}
S.~Roy, A.~S. Sengupta, and P.~Ajith, ``{Effectual template banks for upcoming
  compact binary searches in Advanced-LIGO and Virgo data},'' {\em Phys. Rev.
  D}, vol.~99, no.~2, p.~024048, 2019.

\bibitem{Coogan:2022qxs}
A.~Coogan, T.~D.~P. Edwards, H.~S. Chia, R.~N. George, K.~Freese, C.~Messick,
  C.~N. Setzer, C.~Weniger, and A.~Zimmerman, ``{Efficient gravitational wave
  template bank generation with differentiable waveforms},'' {\em Phys. Rev.
  D}, vol.~106, no.~12, p.~122001, 2022.

\bibitem{Schmidt:2023gzj}
S.~Schmidt, B.~Gadre, and S.~Caudill, ``{Gravitational-wave template banks for
  novel compact binaries},'' {\em Phys. Rev. D}, vol.~109, no.~4, p.~042005,
  2024.

\bibitem{Harry:2016ijz}
I.~Harry, S.~Privitera, A.~Boh\'e, and A.~Buonanno, ``{Searching for
  Gravitational Waves from Compact Binaries with Precessing Spins},'' {\em
  Phys. Rev. D}, vol.~94, no.~2, p.~024012, 2016.

\bibitem{Harry:2017weg}
I.~Harry, J.~Calder\'on~Bustillo, and A.~Nitz, ``{Searching for the full
  symphony of black hole binary mergers},'' {\em Phys. Rev. D}, vol.~97, no.~2,
  p.~023004, 2018.

\bibitem{Allen:2022lqr}
B.~Allen, ``{Performance of random template banks},'' {\em Phys. Rev. D},
  vol.~105, no.~10, p.~102003, 2022.

\bibitem{Allen:2021yuy}
B.~Allen, ``{Optimal template banks},'' {\em Phys. Rev. D}, vol.~104, no.~4,
  p.~042005, 2021.

\bibitem{Apostolatos:1995pj}
T.~A. Apostolatos, ``{Search templates for gravitational waves from precessing,
  inspiraling binaries},'' {\em Phys. Rev. D}, vol.~52, pp.~605--620, 1995.

\bibitem{Apostolatos:1996rf}
T.~A. Apostolatos, ``{Construction of a template family for the detection of
  gravitational waves from coalescing binaries},'' {\em Phys. Rev. D}, vol.~54,
  pp.~2421--2437, 1996.

\bibitem{Buonanno:2005pt}
A.~Buonanno, Y.~Chen, Y.~Pan, H.~Tagoshi, and M.~Vallisneri, ``{Detecting
  gravitational waves from precessing binaries of spinning compact objects. II.
  Search implementation for low-mass binaries},'' {\em Phys. Rev. D}, vol.~72,
  p.~084027, 2005.

\bibitem{McIsaac:2023ijd}
C.~McIsaac, C.~Hoy, and I.~Harry, ``{A search technique to observe precessing
  compact binary mergers in the advanced detector era},'' 3 2023.

\bibitem{Fairhurst:2019vut}
S.~Fairhurst, R.~Green, C.~Hoy, M.~Hannam, and A.~Muir, ``{Two-harmonic
  approximation for gravitational waveforms from precessing binaries},'' {\em
  Phys. Rev. D}, vol.~102, no.~2, p.~024055, 2020.

\bibitem{DalCanton:2014qjd}
T.~Dal~Canton, A.~P. Lundgren, and A.~B. Nielsen, ``{Impact of precession on
  aligned-spin searches for neutron-star\textendash{}black-hole binaries},''
  {\em Phys. Rev. D}, vol.~91, no.~6, p.~062010, 2015.

\bibitem{PhysRevD.102.043005}
A.~Ramos-Buades, S.~Tiwari, M.~Haney, and S.~Husa, ``Impact of eccentricity on
  the gravitational-wave searches for binary black holes: High mass case,''
  {\em Phys. Rev. D}, vol.~102, p.~043005, Aug 2020.

\bibitem{Ramos-Buades:2020eju}
A.~Ramos-Buades, S.~Tiwari, M.~Haney, and S.~Husa, ``{Impact of eccentricity on
  the gravitational wave searches for binary black holes: High mass case},''
  {\em Phys. Rev. D}, vol.~102, no.~4, p.~043005, 2020.

\bibitem{Blanchet:2008je}
L.~Blanchet, G.~Faye, B.~R. Iyer, and S.~Sinha, ``{The Third post-Newtonian
  gravitational wave polarisations and associated spherical harmonic modes for
  inspiralling compact binaries in quasi-circular orbits},'' {\em Class. Quant.
  Grav.}, vol.~25, p.~165003, 2008.
\newblock [Erratum: Class.Quant.Grav. 29, 239501 (2012)].

\bibitem{CalderonBustillo:2015lrt}
J.~Calder\'on~Bustillo, S.~Husa, A.~M. Sintes, and M.~P\"urrer, ``{Impact of
  gravitational radiation higher order modes on single aligned-spin
  gravitational wave searches for binary black holes},'' {\em Phys. Rev. D},
  vol.~93, no.~8, p.~084019, 2016.

\bibitem{Chandra:2022ixv}
K.~Chandra, J.~Calder\'on~Bustillo, A.~Pai, and I.~W. Harry, ``{First
  gravitational-wave search for intermediate-mass black hole mergers with
  higher-order harmonics},'' {\em Phys. Rev. D}, vol.~106, no.~12, p.~123003,
  2022.

\bibitem{LIGOScientific:2019lzm}
R.~Abbott {\em et~al.}, ``{Open data from the first and second observing runs
  of Advanced LIGO and Advanced Virgo},'' {\em SoftwareX}, vol.~13, p.~100658,
  2021.

\bibitem{KAGRA:2023pio}
R.~Abbott {\em et~al.}, ``{Open Data from the Third Observing Run of LIGO,
  Virgo, KAGRA, and GEO},'' {\em Astrophys. J. Suppl.}, vol.~267, no.~2, p.~29,
  2023.

\bibitem{Fairhurst:2019srr}
S.~Fairhurst, R.~Green, M.~Hannam, and C.~Hoy, ``{When will we observe binary
  black holes precessing?},'' {\em Phys. Rev. D}, vol.~102, no.~4, p.~041302,
  2020.

\bibitem{Green:2020ptm}
R.~Green, C.~Hoy, S.~Fairhurst, M.~Hannam, F.~Pannarale, and C.~Thomas,
  ``{Identifying when Precession can be Measured in Gravitational Waveforms},''
  {\em Phys. Rev. D}, vol.~103, no.~12, p.~124023, 2021.

\bibitem{Heinicke:2014ipp}
C.~Heinicke and F.~W. Hehl, ``{Schwarzschild and Kerr Solutions of Einstein's
  Field Equation -- an introduction},'' {\em Int. J. Mod. Phys. D}, vol.~24,
  no.~02, p.~1530006, 2014.

\bibitem{Schmidt:2012rh}
P.~Schmidt, M.~Hannam, and S.~Husa, ``{Towards models of gravitational
  waveforms from generic binaries: A simple approximate mapping between
  precessing and non-precessing inspiral signals},'' {\em Phys. Rev. D},
  vol.~86, p.~104063, 2012.

\bibitem{Thomas:2020uqj}
L.~M. Thomas, P.~Schmidt, and G.~Pratten, ``{New effective precession spin for
  modeling multimodal gravitational waveforms in the strong-field regime},''
  {\em Phys. Rev. D}, vol.~103, no.~8, p.~083022, 2021.

\bibitem{Gerosa:2020aiw}
D.~Gerosa, M.~Mould, D.~Gangardt, P.~Schmidt, G.~Pratten, and L.~M. Thomas,
  ``{A generalized precession parameter $\chi_\mathrm{p}$ to interpret
  gravitational-wave data},'' {\em Phys. Rev. D}, vol.~103, no.~6, p.~064067,
  2021.

\bibitem{Maggiore:2007ulw}
M.~Maggiore, {\em {Gravitational Waves. Vol. 1: Theory and Experiments}}.
\newblock Oxford Master Series in Physics, Oxford University Press, 2007.

\bibitem{Creighton_book}
J.~{Creighton} and W.~{Anderson}, {\em {Gravitational-Wave Physics and
  Astronomy: An Introduction to Theory, Experiment and Data Analysis.}}
\newblock 2011.

\bibitem{Finn:1992xs}
L.~S. Finn and D.~F. Chernoff, ``{Observing binary inspiral in gravitational
  radiation: One interferometer},'' {\em Phys. Rev. D}, vol.~47,
  pp.~2198--2219, 1993.

\bibitem{Jaranowski:1998qm}
P.~Jaranowski, A.~Krolak, and B.~F. Schutz, ``{Data analysis of gravitational -
  wave signals from spinning neutron stars. 1. The Signal and its detection},''
  {\em Phys. Rev. D}, vol.~58, p.~063001, 1998.

\bibitem{Sathyaprakash_2009}
B.~S. Sathyaprakash and B.~F. Schutz, ``Physics, astrophysics and cosmology
  with gravitational waves,'' {\em Living Reviews in Relativity}, vol.~12, mar
  2009.

\bibitem{Capano:2013raa}
C.~Capano, Y.~Pan, and A.~Buonanno, ``{Impact of higher harmonics in searching
  for gravitational waves from nonspinning binary black holes},'' {\em Phys.
  Rev. D}, vol.~89, no.~10, p.~102003, 2014.

\bibitem{Wadekar:2023gea}
D.~Wadekar, J.~Roulet, T.~Venumadhav, A.~K. Mehta, B.~Zackay, J.~Mushkin,
  S.~Olsen, and M.~Zaldarriaga, ``{New black hole mergers in the LIGO-Virgo O3
  data from a gravitational wave search including higher-order harmonics},'' 12
  2023.

\bibitem{Wadekar:2023kym}
D.~Wadekar, T.~Venumadhav, A.~K. Mehta, J.~Roulet, S.~Olsen, J.~Mushkin,
  B.~Zackay, and M.~Zaldarriaga, ``{A new approach to template banks of
  gravitational waves with higher harmonics: reducing matched-filtering cost by
  over an order of magnitude},'' 10 2023.

\bibitem{Wadekar:2024zdq}
D.~Wadekar, T.~Venumadhav, J.~Roulet, A.~K. Mehta, B.~Zackay, J.~Mushkin, and
  M.~Zaldarriaga, ``{A new search pipeline for gravitational waves with
  higher-order modes using mode-by-mode filtering},'' 5 2024.

\bibitem{Allen:2004gu}
B.~Allen, ``{${\chi}^{2}$ time-frequency discriminator for gravitational wave
  detection},'' {\em Phys. Rev. D}, vol.~71, p.~062001, 2005.

\bibitem{Schmidt:2024kxy}
S.~Schmidt and S.~Caudill, ``{A novel signal-consistency test for
  gravitational-wave searches of generic black hole binaries},'' 3 2024.

\bibitem{Cannon:2015gha}
K.~Cannon, C.~Hanna, and J.~Peoples, ``{Likelihood-Ratio Ranking Statistic for
  Compact Binary Coalescence Candidates with Rate Estimation},'' 4 2015.

\bibitem{Owen:1995tm}
B.~J. Owen, ``{Search templates for gravitational waves from inspiraling
  binaries: Choice of template spacing},'' {\em Phys. Rev. D}, vol.~53,
  pp.~6749--6761, 1996.

\bibitem{mbank}
S.~Schmidt, ``\texttt{mbank} - metric bank generation for gravitational waves
  data analysis.'' \url{https://mbank.readthedocs.io/en/latest/}.

\bibitem{owen_metric}
B.~J. Owen, ``Search templates for gravitational waves from inspiraling
  binaries: Choice of template spacing,'' {\em Phys. Rev. D}, vol.~53,
  pp.~6749--6761, Jun 1996.

\bibitem{Messenger:2008ta}
C.~Messenger, R.~Prix, and M.~A. Papa, ``{Random template banks and relaxed
  lattice coverings},'' {\em Phys. Rev. D}, vol.~79, p.~104017, 2009.

\bibitem{Prix:2007ks}
R.~Prix, ``{Template-based searches for gravitational waves: Efficient lattice
  covering of flat parameter spaces},'' {\em Class. Quant. Grav.}, vol.~24,
  pp.~S481--S490, 2007.

\bibitem{Brown:2012qf}
D.~A. Brown, I.~Harry, A.~Lundgren, and A.~H. Nitz, ``{Detecting binary neutron
  star systems with spin in advanced gravitational-wave detectors},'' {\em
  Phys. Rev. D}, vol.~86, p.~084017, 2012.

\bibitem{Hanna:2022zpk}
C.~Hanna {\em et~al.}, ``{Binary tree approach to template placement for
  searches for gravitational waves from compact binary mergers},'' {\em Phys.
  Rev. D}, vol.~108, no.~4, p.~042003, 2023.

\bibitem{norm_flow}
G.~Papamakarios, E.~Nalisnick, D.~J. Rezende, S.~Mohamed, and
  B.~Lakshminarayanan, ``Normalizing flows for probabilistic modeling and
  inference,'' 2019.

\bibitem{nflows_paper}
C.~Durkan, A.~Bekasov, I.~Murray, and G.~Papamakarios, ``Neural spline flows,''
  in {\em Advances in Neural Information Processing Systems} (H.~Wallach,
  H.~Larochelle, A.~Beygelzimer, F.~d\textquotesingle Alch\'{e}-Buc, E.~Fox,
  and R.~Garnett, eds.), vol.~32, Curran Associates, Inc., 2019.

\bibitem{Kobyzev_2021}
I.~Kobyzev, S.~J. Prince, and M.~A. Brubaker, ``Normalizing flows: An
  introduction and review of current methods,'' {\em {IEEE} Transactions on
  Pattern Analysis and Machine Intelligence}, vol.~43, pp.~3964--3979, nov
  2021.

\bibitem{Khan:2015jqa}
S.~Khan, S.~Husa, M.~Hannam, F.~Ohme, M.~P\"urrer, X.~Jim\'enez~Forteza, and
  A.~Boh\'e, ``{Frequency-domain gravitational waves from nonprecessing
  black-hole binaries. II. A phenomenological model for the advanced detector
  era},'' {\em Phys. Rev. D}, vol.~93, no.~4, p.~044007, 2016.

\bibitem{O4_PSDs}
R.~Abbott {\em et~al.}, ``Noise curves used for simulations in the update of
  the observing scenarios paper.''

\bibitem{Cannon:2010qh}
K.~Cannon, A.~Chapman, C.~Hanna, D.~Keppel, A.~C. Searle, and A.~J. Weinstein,
  ``{Singular value decomposition applied to compact binary coalescence
  gravitational-wave signals},'' {\em Phys. Rev. D}, vol.~82, p.~044025, 2010.

\bibitem{Cannon:2011vi}
K.~Cannon {\em et~al.}, ``{Toward Early-Warning Detection of Gravitational
  Waves from Compact Binary Coalescence},'' {\em Astrophys. J.}, vol.~748,
  p.~136, 2012.

\bibitem{Sakon:2022ibh}
S.~Sakon {\em et~al.}, ``{Template bank for compact binary mergers in the
  fourth observing run of Advanced LIGO, Advanced Virgo, and KAGRA},'' 11 2022.

\bibitem{Morisaki:2020oqk}
S.~Morisaki and V.~Raymond, ``{Rapid Parameter Estimation of Gravitational
  Waves from Binary Neutron Star Coalescence using Focused Reduced Order
  Quadrature},'' {\em Phys. Rev. D}, vol.~102, no.~10, p.~104020, 2020.

\bibitem{Vallisneri:2014vxa}
M.~Vallisneri, J.~Kanner, R.~Williams, A.~Weinstein, and B.~Stephens, ``{The
  LIGO Open Science Center},'' {\em J. Phys. Conf. Ser.}, vol.~610, no.~1,
  p.~012021, 2015.

\bibitem{Tiwari:2017ndi}
V.~Tiwari, ``{Estimation of the Sensitive Volume for Gravitational-wave Source
  Populations Using Weighted Monte Carlo Integration},'' {\em Class. Quant.
  Grav.}, vol.~35, no.~14, p.~145009, 2018.

\end{thebibliography}
	\bibliographystyle{ieeetr}

\end{document}